\documentclass[preprint,superscriptaddress]{revtex4-1}
\usepackage{xcolor}
\usepackage{tabu}
\usepackage{titlesec}
\usepackage{graphicx,epsfig,epstopdf}
\usepackage{bm}
\usepackage{epstopdf}
\usepackage{amssymb,amsmath,amsfonts,latexsym}
\usepackage{dcolumn}
\usepackage{bm}
\usepackage{hyperref}
\usepackage[utf8]{inputenc}
\newcommand{\bea}{\begin{eqnarray}}
\newcommand{\ena}{\end{eqnarray}}
\newcommand{\be}{\begin{equation}}
\newcommand{\en}{\end{equation}}
\newcommand{\nn}{\nonumber\\}
\newcommand{\no}{\nonumber}
\newcommand{\ed}{\end{document}}

\newcommand{\ord}{\mathcal{O}}
\usepackage{breqn}             
\makeatletter                  
\let\cat@comma@active\@empty   
\makeatother                   

\begin{document}
\title{Rare $b \to d$ decays in covariant confined quark model}
\author{N. R. Soni}
\email{nrsoni-apphy@msubaroda.ac.in}
\affiliation{Department of Physics, Faculty of Science, \\
The Maharaja Sayajirao University of Baroda, Vadodara 390002, Gujarat, {\it INDIA}}

\author{A. Issadykov}
\email{issadykov@jinr.ru}
\affiliation{Bogoliubov Laboratory of Theoretical Physics, \\
Joint Institute for Nuclear Research, 141980 Dubna, \textit{RUSSIA}}
\affiliation{The Institute of Nuclear Physics, \\
Ministry of Energy of the Republic of Kazakhstan, 050032 Almaty,  {\it KAZAKHSTAN}}
\affiliation{Al-Farabi Kazakh National University, 71 al-Farabi, 050038 Almaty, {\it KAZAKHSTAN}}

\author{A. N. Gadaria}
\email{angadaria-apphy@msubaroda.ac.in}
\affiliation{Applied Physics Department, Faculty of Technology and Engineering,  \\
The Maharaja Sayajirao University of Baroda, Vadodara 390001, Gujarat, {\it INDIA}}

\author{J. J. Patel}
\email{jjpatel-apphy@msubaroda.ac.in}
\affiliation{Department of Physics, Faculty of Science, \\
The Maharaja Sayajirao University of Baroda, Vadodara 390002, Gujarat, {\it INDIA}}
\affiliation{Applied Physics Department, Faculty of Technology and Engineering, \\
The Maharaja Sayajirao University of Baroda, Vadodara 390001, Gujarat, {\it INDIA}}

\author{J. N. Pandya}
\email{jnpandya-apphy@msubaroda.ac.in}
\affiliation{Applied Physics Department, Faculty of Technology and Engineering, \\
The Maharaja Sayajirao University of Baroda, Vadodara 390001, Gujarat, {\it INDIA}}

\date{\today}

\begin{abstract}
In this article, we study the rare decays corresponding to $b \to d$ transition in the framework of the covariant confined quark model. The transition form factors for the channels $B^{+(0)} \to (\pi^{+(0)}, \rho^{+(0)},\omega)$ and $B_s^0 \to K^{(*)0}$ are computed in the entire dynamical range of momentum transfer squared. Using the form factors, we compute the branching fractions of the rare decays and our results are found to be matching well with the experimental data.
We also compute the ratios of the branching fractions of the  $b \to s$ to $b \to d$ rare decays using the inputs from previous papers on $b \to s \ell^+\ell^-$ using this model.
Further, using the form factors, model dependent and independent parameters, we also compute different other physical observables such as forward backward asymmetry, longitudinal polarization and angular observables in the entire $q^2$ range as well as in $q^2$ bins [0.1 -- 0.98] GeV$^2$ and [1.1 -- 6] GeV$^2$. We also compare our findings with different theoretical predictions.
\end{abstract}

\maketitle

\section{Introduction}
The flavor changing neutral current decays $b \to s$ have served as important probe for testing the standard model as well as in search of physics beyond the standard model in the light of the experimental results.
Experimentally, anomalies have been reported in the channels $B \to K^{(*)} \ell^+ \ell^-$ and $B \to D^{(*)} \ell^+ \nu_\ell$ \cite{Aaij:2013qta,Aaij:2017vbb,Aaij:2014ora,Huschle:2015rga,Sato:2016svk,Aaij:2015yra,Aaij:2015yra} that hint towards the violation of lepton flavor universality as their results deviate from the standard model predictions \cite{Ali:1999mm,Beneke:2001at,Chen:2001ri,Ali:2006ew,Egede:2008uy,Bobeth:2008ij,Altmannshofer:2008dz}.
In the SM, the $B \to K^{(*)} \ell^+ \ell^-$ decays occur at the electroweak loop level and are generally suppressed at the tree level.
Similarly, $b \to d \ell^+\ell^-$ decays can also serve as important probe as they also follow the same flavor changing neutral current (FCNC) at quark level \cite{Hurth:2010tk,Blake:2016olu}.
Though there exists rich data for $b \to s \ell^+ \ell^-$ induced processes, the $b \to d \ell^+ \ell^-$ counterpart of the weak decay has not caught much attention perhaps due to low branching fraction.
Within the standard model, the branching fractions are suppressed by a factor of $|V_{td}/V_{ts}|^2$ compared to the $b \to s$ transitions.
Experimentally, these transitions have been observed by LHCb collaboration in the channel $B^+ \to \pi^+ \mu^+ \mu^-$ \cite{Aaij:2015nea} and in $\Lambda_b^0 \to p \pi^- \mu^+ \mu^-$ \cite{Aaij:2017ewm} and recently, they have also reported the evidence for the $B_s^0 \to \bar{K}^{*0} \mu^+ \mu^-$ with 3.4 standard deviation significance \cite{Aaij:2018jhg}. Their results read
\begin{eqnarray}
\mathcal{B} (B^+ \to \pi^+ \mu^+ \mu^-) & = & (1.83 \pm 0.24 \pm 0.05) \times 10^{-8} ~\text{\cite{Aaij:2015nea}}\nn
\mathcal{B} (B_s^0 \to \bar{K}^*(892)^0 \mu^+ \mu^-) & = & (2.9 \pm 1.0 \pm 0.3 \pm 0.3) \times 10^{-8} ~\text{\cite{Aaij:2018jhg}} \nonumber
\end{eqnarray}
Further, the Belle collaboration has also measured the branching fractions for the decay channel $B^+ \to \pi^+ \pi^- \ell^+ \ell^-$  very recently \cite{Beleno:2020gzt}.
They have also performed the search for different channels corresponding to rare $b \to (s, d) \nu \bar{\nu}$ transitions \cite{Chen:2007zk,Lutz:2013ftz,Grygier:2017tzo}.

There are several theoretical studies dedicated to the investigation of $b \to d$ transitions.
Form factors for $B \to \pi, \rho , K^{(*)}$ and $B_s \to K^{(*)}$ decays were calculated in light cone sum rules  ~\cite{Lu:2018cfc,Wu:2006rd,Khodjamirian:2017fxg,Gubernari:2018wyi,Ball:2004rg,Hambrock:2015wka,Straub:2015ica,Cheng:2018ouz}.
Very recently, the $B \to V$ form factors are computed precisely using the soft-collinear effective field theory with light cone sum rules \cite{Gao:2019lta}.
Lattice calculations of $B$ semileptonic form factors can be found in the Refs. \cite{Okamoto:2004xg,Dalgic:2006dt,Bailey:2008wp}.
Also, $B_{(s)} \to K^*$ and $B_s \to \phi$ form factors were reported by Horgan \textit{et al}., in lattice QCD~\cite{Horgan:2013hoa,Horgan:2015vla}. The lattice QCD calculations of $B \to \pi \ell^+\ell^-$ form factors and branching fractions were initially reported by Fermilab Lattice and MILC Collaborations \cite{Du:2015tda,Bailey:2015nbd,Lattice:2015tia}.
Branching fractions and forward-backward asymmetry for $B^+ \to (\pi^+,\rho^+) \ell^+ \ell^-$ were studied in the R-parity violating supersymmetric standard model~\cite{Wang:2007sp} and in non-universal Z' model~\cite{Nayek:2018rcq}.
Moreover predictions of decay rates and angular observables for  $B \to \rho\ell^+ \ell^-$ and $B_s \to K^{(*)} \ell^+ \ell^-$ were given in ref. ~\cite{Kindra:2018ayz} within the standard model and also in recent ref. \cite{Jin:2020jtu} using the perturbative QCD factorization approach with lattice input.
The angular observables for $B_s \to \bar{K}^*\ell\ell$ are also computed in the non-universal $Z^\prime$ model \cite{Alok:2019xub}.
$B_{(s)}$ transition form factors are also computed in the perturbative QCD approach \cite{Li:2009tx,Wang:2012ab}.
The heavy to light form factors were also computed in light cone quark model using soft collinear effective field theory \cite{Lu:2007sg}.
The form factors as well as branching fractions were computed using the relativistic quark model based on quasi potential approach \cite{Faustov:2013ima,Faustov:2013pca,Faustov:2014zva}, constituent quark model \cite{Melikhov:2000yu} and light front quark model \cite{Choi:2010zb,Verma:2011yw,Chang:2019mmh}.

Further, it has been observed that the branching fraction ratios of $b \to s \ell^+ \ell^-$ and $b \to d \ell^+ \ell^-$ could be the probe for new physics beyond the standard model as it would provide stringent tests of the flavor structure of the underlying interactions as well as allow one to study the hypothesis of minimal flavor violation \cite{DAmbrosio:2002vsn,Albrecht:2018vsa}.
Also this ratio would provide the determination of the ratio of Cabibbo-Kobayashi-Maskawa (CKM) matrix
$|V_{td}/V_{ts}|$.
Experimentally, this ratio was observed by LHCb collaboration \cite{LHCb:2012de} in the channel $B^+ \to \pi^+$ and $B^+ \to K^+$.
Also the ratio of branching fractions was observed for the channels $B_s^0 \to \bar{K}^{*}(892)^0$ and $B^0 \to K^{*}(892)^0$ as well as in the two body decays of $B_s$ meson \cite{Aaij:2015mea}.
Couple of notable results read,
\begin{eqnarray}
\frac{\mathcal{B} (B^+ \to \pi^+ \mu^+ \mu^-)}{\mathcal{B} (B^+ \to K^+ \mu^+ \mu^-)} & = & (5.3 \pm 1.4 \pm 0.1) \% ~\text{\cite{LHCb:2012de}} \nn
\frac{\mathcal{B}(B_s^0\to \bar{K}^*(892)^0\mu^+ \mu^-)}{\mathcal{B}(\bar{B}^0\to \bar{K}^*(892)^0\mu^+ \mu^-)} & = & (3.3 \pm 1.1 \pm 0.3 \pm 0.2) \% ~\text{\cite{Aaij:2018jhg}}\nonumber
\end{eqnarray}
In this paper, we compute the rare decays corresponding to $b \to d$ transition involving the various  channels $B^{+(0)} \to (\pi^{+(0)},\rho^{+(0)}, \omega) \ell^+ \ell^-$ and $B_s^0 \to \bar{K}^{(*)0} \ell^+ \ell^-$ for $\ell = e, \mu$ and $\tau$ within the standard model framework of covariant confined quark model (CCQM).
The form factors are computed in the entire physical range of momentum transfer employing the covariant confined quark model with built-in infrared confinement \cite{Efimov:1993,Branz:2009cd,Ivanov:2011aa,Gutsche:2012ze}. These transition form factors are then used for computation of various physical observables such as branching fractions, forward backward asymmetry, longitudinal polarizations and also various angular observables. We further provide the ratios of the branching fractions corresponding to the rare decay of $b \to s$ and $b \to d$.
We also present brief comparison of our results with few other theoretical predictions and available experimental data.

The rest of the paper is organised in the following way.
After the brief introduction of the subject with literature survey, in Sec. \ref{sec:framework}, we introduce the effective Hamiltonian framework for studying the rare decays.
Further, we briefly introduce the CCQM for computations of transition form factors.
Using the effective Hamiltonian and transition form factors, we compute the branching fractions, forward-backward asymmetry, longitudinal and transverse polarizations and angular observables.
In Sec. \ref{sec:result}, we provide all the numerical results in comparison with theoretical predictions and available experimental data.
Finally, in Sec. \ref{sec:conclusion}, we summarize and conclude the presented work.

\section{Theoretical Framework}
\label{sec:framework}
Within the Standard Model (SM), the effective Hamiltonian for the $b\to d\ell^+\ell^-$ decay can be written in terms of operator product expansion as \cite{Buras:1994dj,Kruger:1996dt,Buchalla:1995vs}
\begin{eqnarray}
\mathcal{H}^{SM}_{eff}=-\frac{4G_F}{\sqrt{2}}V^*_{td}V_{tb}\left\{\sum^{10}_{i=1}C_{i}(\mu)\mathcal{O}_i(\mu)
+\lambda_u\sum^{2}_{i=1}C_{i}(\mu)[\mathcal{O}_{i}(\mu)-\mathcal{O}^{(u)}_i(\mu)]\right\},
\label{eq:hamiltonian}
\end{eqnarray}
where $\lambda_u\equiv\frac{V_{ub}^{*}V_{ud}}{V^{*}_{tb}V_{td}}$.

In the above equation, $C_i$ are the Wilson coefficients and the set of local operators $\mathcal{O}_i$ obtained within the SM for~$b \to d \ell^+ \ell^-$ transition can be written in standard form as \cite{Kruger:1996dt,Buchalla:1995vs}
\begin{eqnarray}
\begin{array}{ll}
\ord_{1}^{u}     =  (\bar{d}_{a_1}\gamma^\mu P_L u_{a_2})
              (\bar{u}_{a_2}\gamma_\mu P_L b_{a_1}),                   &
\ord_{2}^{u}     =  (\bar{d}\gamma^\mu P_L u)  (\bar{u}\gamma_\mu P_L b),
\\[2ex]
\ord_{1}     =  (\bar{d}_{a_1}\gamma^\mu P_L c_{a_2})
              (\bar{c}_{a_2}\gamma_\mu P_L b_{a_1}),                   &
\ord_{2}     =  (\bar{d}\gamma^\mu P_L c)  (\bar{c}\gamma_\mu P_L b),
\\[2ex]
\ord_3     =  (\bar{d}\gamma^\mu P_L  b) \sum_q(\bar{q}\gamma_\mu P_L q),  &
\ord_4     =  (\bar{d}_{a_1}\gamma^\mu P_L  b_{a_2})
              \sum_q (\bar{q}_{a_2}\gamma_\mu P_L q_{a_1}),
\\[2ex]
\ord_5     =  (\bar{d}\gamma^\mu P_L b)
              \sum_q(\bar{q}\gamma_\mu P_R q),            &
\ord_6     =  (\bar{d}_{a_1}\gamma^\mu P_L b_{a_2 })
              \sum_q  (\bar{q}_{a_2} \gamma_\mu P_R q_{a_1}),
\\[2ex]
\ord_7     =  \frac{e}{16\pi^2} \bar m_b\,
              (\bar{d} \sigma^{\mu\nu} P_R b) F_{\mu\nu},       &
\ord_8    =  \frac{g}{16\pi^2} \bar m_b\,
              (\bar{d}_{a_1} \sigma^{\mu\nu} P_R {\bf T}_{a_1a_2} b_{a_2})
              {\bf G}_{\mu\nu},
\\[2ex]
\ord_9     = \frac{e^2}{16\pi^2}
             (\bar{d} \gamma^\mu P_L b) (\bar\ell\gamma_\mu \ell),     &
\ord_{10}  = \frac{e^2}{16\pi^2}
             (\bar{d} \gamma^\mu P_L b)  (\bar\ell\gamma_\mu\gamma_5 \ell),
\end{array}
\label{eq:operators}
\end{eqnarray}
\begin{table}
\caption{Masses, total widths and dilepton decay widths of vector resonance states \cite{Tanabashi:2018oca}}
\begin{tabular*}{\textwidth}{@{\extracolsep{\fill}}cccc@{}}
\hline\hline
State & Mass (MeV) & $\Gamma_V$ (MeV) & $\mathcal{B}(V \to \ell^+ \ell^-)$\\
\hline
$\rho$ 			& 775.26		& 147.8							& $4.63 \times 10^{-5}$\\
$\omega$		& 785.65		& 8.49							& $7.38 \times 10^{-5}$\\
$\phi$ 			& 1.019			& 4.249							& $2.94 \times 10^{-4}$\\
$J/\psi$			& 3096.900	& 92.9 $\times 10^{-3}$	& $5.96 \times 10^{-2}$\\
$\psi (2S)$	& 3686.10		& 294 $\times 10^{-3}$	& $7.96 \times 10^{-3}$\\
\hline\hline
\end{tabular*}
\label{tab:resonance}
\end{table}
\begin{table*}[!htbp]
\caption{Values of the input parameters and SM Wilson coefficients \cite{Jin:2020jtu,Buchalla:1995vs}.}
\begin{tabular*}{\textwidth}{@{\extracolsep{\fill}}cccccccccc@{}}
\hline\hline
 $m_W$ &  $\sin^2\theta_W $ &  $\alpha(M_Z)$ &
$\bar m_c$ &  $\bar m_b$  &  $\bar m_t$ & $ \lambda_d$  & $ \lambda_s$&\\
\hline
 $80.41$~GeV & $0.2313$ & $1/128.94$ & $1.27$~GeV & $4.68$~GeV & $173.3$~GeV &
 0.00825 &0.0401 & \\
\hline\hline
$C_1$ &  $C_2$ &  $C_3(\%)$ &  $C_4(\%)$ & $C_5(\%)$ &  $C_6(\%)$ &  $C_7$ & $C_8$ &
$C_9$ &  $C_{10}$ \\
\hline
$-0.175$ & $1.076$ &  $1.258$ & $-3.279$ & $1.112$ & $-3.634$ & $-0.302$ & $-0.148$ &
4.232 & $-4.410$\\
\hline\hline
\end{tabular*}
\label{tab:input}
\end{table*}
where ${\bf G}_{\mu\nu}$ and $F_{\mu\nu}$ are the gluon and photon
field strengths, respectively; ${\bf T}_{a_1a_2}$ are the generators of
the $SU(3)$ color group; $a_1$ and $a_2$ denote color indices
(they are omitted in the color-singlet currents).
The chirality projection operators are
$P_{L,R} = (1 \mp \gamma_5)/2$ and $\mu$ is a renormalization scale.
$\ord_{1,2}$ are current-current operators,
$\ord_{3-6}$ are QCD penguin operators,  $\ord_{7,8}$ are dipole operators, and $\ord_{9,10}$ are semileptonic electroweak penguin operators. We denote the QCD quark masses by the bar symbol
to distinguish them from the constituent quark masses used in the model.
The matrix  element for $b\to d\ell^+\ell^-$ can be written as \cite{Buras:1994dj,Kruger:1996dt}
\begin{eqnarray}
\mathcal{M}(b\to d
\ell^+\ell^-)&=&\frac{G_F\alpha}{\sqrt{2}\pi}V^{*}_{tb}V_{td}\Biggl\{C^{\mathrm{eff}}_9
(\bar{d}\gamma_\mu P_{L} b)(\bar{\ell}\gamma^\mu
\ell)+C_{10}(\bar{d}\gamma_\mu
P_{L} b)(\bar{\ell}\gamma^\mu\gamma_5\ell)\nonumber\\
&&- \frac{2\bar m_b}{q^2}
C^{\mathrm{eff}}_7\left(\bar{d} i \sigma_{\mu\nu}q_\nu P_{R} b\right)
(\bar{\ell}\gamma^\mu \ell)\Biggl\},\label{quarkM}
\end{eqnarray}
where the effective Wilson coefficients are given by \cite{Chen:2001zc,Wang:2012ab}
\begin{eqnarray}
C^{\mathrm{eff}}_7 (\mu) = C_7(\mu) + i \alpha_s \left\{\frac29 \left(\frac{\alpha_s (m_W)}{\alpha_s (\mu)}\right)^{14/23} [G_I(x_t) - 0.1687] - 0.03 C_2 (\mu)\right\}
\end{eqnarray}
with $x_t = m_t^2/m_W^2$ and
\begin{eqnarray}
G_I (x_t) = \frac{x_t (x_t^2 - 5 x_t - 2)}{8 (x_t - 1)^3} + \frac{3 x_t^2 \mathrm{ln}~x_t}{4 (x_t - 1)^4}.\no
\end{eqnarray}

Further, $C^{\mathrm{eff}}_{9}(\mu)$ contains the corrections of  four-quark operators $\mathcal{O}_{1-6}$ and
$\mathcal{O}^{u}_{1,2}$ in Eq. (\ref{eq:hamiltonian}), which  can be
written as  \cite{Deshpande:1988bd,Jezabek:1988ja,Lim:1988yu,Misiak:1992bc,ODonnell:1991cdx,Ali:1991is,Bobeth:1999mk,Chen:2001zc,Wang:2012ab}
\begin{eqnarray}
C^{\mathrm{eff}}_9(\mu)=\xi_1+\lambda^{*}_u\xi_2,
\end{eqnarray}
with
\begin{eqnarray}
\xi_1 & = & C_9 + C_0 h^{\mathrm{eff}} (\hat{m}_c, \hat{s}) - \frac12 h(1,\hat{s}) (4 C_3 + 4 C_4 +3 C_5 + C_6) \nn
&-& \frac12 h(0, \hat{s}) (C_3 + 3 C_4) + \frac29 (3 C_3 + C_4 + 3 C_5 + C_6)\\
\xi_2 & = & \Big[h^{\mathrm{eff}} (\hat{m}_c, \hat{s}) - h^{\mathrm{eff}} (\hat{m}_u, \hat{s})\Big] (3C_1+C_2)
\end{eqnarray}
where $C_0\equiv 3 C_1 + C_2 + 3 C_3 + C_4+ 3 C_5 + C_6$.
Here, the charm-loop function can be written as
\bea
h(\hat m_q,  \hat{s}) & = & - \frac{8}{9}\ln\hat m_q +
\frac{8}{27} + \frac{4}{9} x
- \frac{2}{9} (2+x) |1-x|^{1/2} \left\{
\begin{array}{ll}
\left( \ln\left| \frac{\sqrt{1-x} + 1}{\sqrt{1-x} - 1}\right| - i\pi
\right), &
\mbox{for } x \equiv \frac{4 \hat m_q^2}{\hat{s}} < 1, \nonumber \\
 & \\
2 \arctan \frac{1}{\sqrt{x-1}}, & \mbox{for } x \equiv \frac
{4 \hat m_q^2}{\hat {s}} > 1,
\end{array}
\right.\nonumber
\ena
and
\bea
h(0, \hat{s}) & = & \frac{8}{27} - \frac{4}{9} \ln \hat{s} + \frac{4}{9} i\pi,
\nonumber
\ena

further the functions,
\begin{eqnarray}
h^{\mathrm{eff}} (\hat{m}_c, \hat{s}) & = & h(\hat m_c,  \hat{s})  +  \frac{3 \pi}{\alpha^2 C_0} \sum_{V = J/\psi, \psi(2S), ...} \frac{m_V \mathcal{B}(V \to \ell^+ \ell^-) \Gamma_V}{m_V^2 - q^2 - i m_V \Gamma_V} , \nn
h^{\mathrm{eff}} (\hat{m}_u, \hat{s}) & = & h(\hat m_u,  \hat{s}) + \frac{3 \pi}{\alpha^2 C_0} \sum_{V = \rho^0, \omega, \phi} \frac{m_V \mathcal{B}(V \to \ell^+ \ell^-) \Gamma_V}{m_V^2 - q^2 - i m_V \Gamma_V}
\label{eq:resonance}
\end{eqnarray}
where $\hat m_q=\bar m_q/m_1$, $\hat s=q^2/m_1^2$.
The nonresonant contribution is computed by ignoring the terms containing the vector resonances in Eq. (\ref{eq:resonance}).
The masses, total decay widths and dilepton branching fractions are inputs from PDG data \cite{Tanabashi:2018oca} and are listed in Tab. \ref{tab:resonance}.
We will use the value of $\mu=\bar m_{b\,\rm pole}$ for the renormalization scale.

We employ the next-to-leading order SM Wilson coefficients from Ref.~\cite{Jin:2020jtu,Buchalla:1995vs} and they are computed at the renormalization scale $\mu_b= 4.8$~GeV.
The values of the model independent input parameters and the Wilson coefficients are listed in Table~\ref{tab:input}.

Now, we define the form factors for $B_{(s)} \to (\pi, \bar{K}^0) \ell^+\ell^-$ in the formalism of CCQM by
\begin{eqnarray}\
\langle \pi, \bar{K}^0 (p_2) &|& \bar{d} O^\mu b ~|~ B_{(s)} (p_1) \rangle \nn
&=& N_c g_{B_{(s)}} g_{\pi (K)} \int \frac{d^4 k}{(2\pi)^4 i} \tilde{\phi}_{B{(s)}} (-(k + w_{13} p_1)^2) \tilde{\phi}_{\pi (K)}(-(k + w_{23} p_2)^2) \cr && \times \mathrm{tr}[O^\mu S_1(k + p_1) \gamma^5 S_3(k) \gamma^5 S_2(k + p_2)] \cr
&=& F_+(q^2) P^{\mu} +  F_-(q^2) q^{\mu} ~, \nn
\langle \pi, \bar{K}^0 (p_2) &|& \bar{d} \sigma^{\mu\nu} (1 - \gamma^5) b ~|~ B_{(s)} (p_1) \rangle \nn & = & N_c g_{B_{(s)}} g_{\pi (K)} \int \frac{d^4 k}{(2\pi)^4 i} \tilde{\phi}_{B{(s)}} (-(k + w_{13} p_1)^2) \tilde{\phi}_{\pi (K)}(-(k + w_{23} p_2)^2) \cr && \times \mathrm{tr}[\sigma^{\mu\nu} (1 - \gamma^5) S_1(k + p_1) \gamma^5 S_3(k) \gamma^5 S_2(k + p_2)] \nn
& = & \frac{i F_T(q^2)}{m_1 + m_2} (P^\mu q^\nu - P^\nu q^\mu + i \varepsilon^{\mu\nu Pq}).
\label{eq:ff_PP}
\end{eqnarray}
and the form factors for $B_{(s)} \to (\rho, \omega^0, \bar{K}^*(892)^0) \ell^+ \ell^-$ can be written as
\begin{eqnarray}
\langle \rho, \omega^0, \bar{K}^{*0} (p_2, \epsilon)&|& \bar{d} O^\mu b ~|~ B_{(s)} (p_1) \rangle  \nn &=& N_c g_{B_{(s)}} g_{\rho, (K^{*0})} \int \frac{d^4 k}{(2\pi)^4 i} \tilde{\phi}_{B_{(s)}} (-(k + w_{13} p_1)^2) \tilde{\phi}_{\rho, (K^{*0})}(-(k + w_{23} p_2)^2) \cr && \times \mathrm{tr}[O^\mu S_1(k + p_1) \gamma^5 S_3(k) \not\!{\epsilon}_{\nu}^\dag S_2(k + p_2)]
\cr & =&\frac{\epsilon_{\nu}^{\dag}}{m_1 + m_2} \left[ -g^{\mu\nu} P\cdot q A_0(q^2) + P^{\mu} P^{\nu}  A_+(q^2) + q^{\mu} P^{\nu}  A_-(q^2) \right. \cr && \left. + i\varepsilon^{\mu\nu\alpha\beta} P_{\alpha} q_{\beta}V(q^2) \right ] ~,\nn
\langle \rho, \omega^0, \bar{K}^{*0} (p_2, \epsilon)&|& \bar{d} \sigma^{\mu\nu} q_\nu (1 + \gamma^5) b ~|~ B_{(s)} (p_1) \rangle \nn & = & N_c g_{B_{(s)}} g_{\rho, (K^{*0})} \int \frac{d^4 k}{(2\pi)^4 i} \tilde{\phi}_{B_{(s)}} (-(k + w_{13} p_1)^2) \tilde{\phi}_{\rho, (K^{*0})}(-(k + w_{23} p_2)^2) \cr && \times \mathrm{tr}[\sigma^{\mu\nu} q_\nu (1 + \gamma^5) S_1(k + p_1) \gamma^5 S_3(k) \not\!{\epsilon}_{\nu}^\dag S_2(k + p_2)] \nn
& = & \epsilon_\nu^\dagger (- (g^{\mu\nu} - q^\mu q^\nu/q^2) P\cdot q a_0(q^2) \nn & + & (P^\mu P^\nu - q^\mu P^\nu P \cdot q/q^2) a_+(q^2) + i \varepsilon^{\mu\nu\alpha\beta} P_\alpha q_\beta g(q^2)).
\label{eq:ff_PV}
\end{eqnarray}

In the above equations, $P = p_1 + p_2$ and $q = p_1 - p_2$ with $p_1$ and $p_2$ to be the momenta of $B_{(s)}$ of mass $m_1$ and daughter meson of mass $m_2$, respectively. Also, $\epsilon$ is the polarization vector of the daughter meson and $O^\mu = \gamma^\mu (1 - \gamma^5)$ is the weak Dirac matrix. 
The on-shell condition also requires that $p_1^2 = m_1^2 = m_{B_{(s)}}^2$ and $p_2^2 = m_2^2 = m_{P/V}^2$ with $P = \pi^+, \pi^0, \bar{K}^0$ and $V = \rho^+, \rho^0, \omega, \bar{K}^*(892)^0$.

The form factors appearing in the above equations are computed in the framework of CCQM which is the effective quantum field theoretical approach for hadronic interaction with constituent quark \cite{Efimov:1993,Branz:2009cd,Ivanov:2011aa,Gutsche:2012ze}.
Here we point out key features of the model relevant to the present study.

The Lagrangian describing the interaction between the hadron with the constituent quark can be written as \cite{Ivanov:1999ic},
\begin{eqnarray}
\mathcal{L}_{int}  &=&  g_M M(x) \int dx_1 dx_2 F_M(x;x_1,x_2) \bar{q}_2(x_2) \Gamma_M q_1(x_1)   H.c.
\label{eq:lagrangian}
\end{eqnarray}
The interaction Lagrangian is written here for meson field only and it can be generalised for baryons and multiquark states as well.
In the above equation, the Dirac matrix $\Gamma_M = I, \gamma^5, \gamma_\mu$ for scalar, pseudoscalar and vector mesons respectively. $g_M$ is the strength corresponding to the coupling between the hadron and its constituent and is determined using the Compositeness conditions. The Compositeness condition \cite{Salam:1962,Weinberg:1962} requires the renormalization constant for the bare state to composite meson state to be equal to zero. Mathematically, this can be achieved by renormalization of self energy Feynman diagram. The Compositeness condition essentially guarantees that the final hadronic state does not contain any bare quark as well as avoids the double counting of hadronic degree of freedom.
In Eq. (\ref{eq:lagrangian}), $F_M$ corresponds to the vertex function which is related to scalar part of Bethe Salpeter equation of the form $F_M(x;x_1,x_2) = \delta (x - w_1 x_1 - w_2 x_2) \times \Phi_M ((x_1 - x_2)^2)$ with $w_i = m_{q_i}/(m_{q_1} + m_{q_2})$.
The vertex function also describes the distribution of quark within the hadron and hence depends on the effective physical size of the hadron.
We choose the vertex function to be of the Gaussian form $\tilde{\Phi}_M(-k^2) = exp (k^2/\Lambda_M^2)$ considering the fact that it should not include any ultraviolet divergence in the quark loop diagram as well as its Fourier transform does have appropriate fall-off behaviour in the Euclidean region.
Here, the model parameter $\Lambda_M$ characterizes the physical size of the meson.
Note that in Eq. (\ref{eq:ff_PP}) and (\ref{eq:ff_PV}) we take $w_{ij} = m_{q_j}/(m_{q_i} + m_{q_j})$ as there are three quarks involved in the semileptonic transition form factors.
The Feynman diagram for hadronic transitions can be drawn using the convolution of quark propagator and vertex function. The loop integrals are evaluated using the Fock - Schwinger representation of the quark propagators ($S_{1,2,3}$ in Eq. (\ref{eq:ff_PP}) and (\ref{eq:ff_PV})). Finally, the universal infrared cutoff parameter $\lambda$ is introduced in computation which removes possible threshold in the quark loop diagram which also guarantees the quark confinement within the hadrons \cite{Branz:2009cd}. We take $\lambda$ to be the same for all the physical processes.
\begin{table*}[!htbp]
\caption{CCQM model parameters: quark masses, meson size parameters and infra-red cut-off parameter (all in GeV)}
\begin{tabular*}{\textwidth}{@{\extracolsep{\fill}}cccc|ccccccc|c@{}}
\hline\hline
$m_{u/d}$        &      $m_s$        &      $m_c $      &     $m_b$  & $\Lambda_{B}$ &$\Lambda_{B_s}$ & $\Lambda_\pi$ & $\Lambda_{\bar K^0}$ &$\Lambda_{\rho}$ &   $\Lambda_{\bar{K}^*(892)^0}$ &  $\Lambda_\omega$ & $\lambda$\\
\hline
0.241 & 0.428 & 1.67 & 5.05 & 1.963 & 2.05 & 0.871 & 1.014 & 0.610  & 0.81 & 0.488 & 0.181\\
\hline\hline
\end{tabular*}
\label{tab:ccqm_parameters}
\end{table*}

\begin{figure*}[!htbp]
\centering
\includegraphics[width=0.45\textwidth]{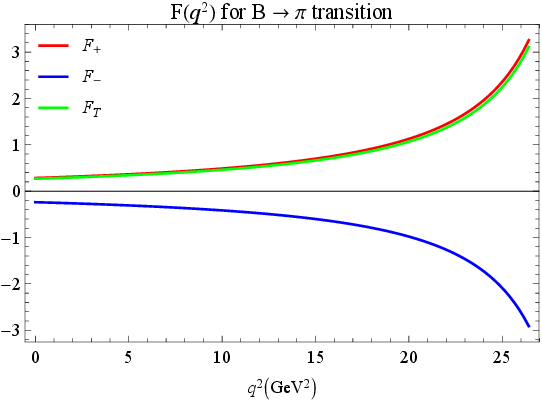}
\hfill\includegraphics[width=0.45\textwidth]{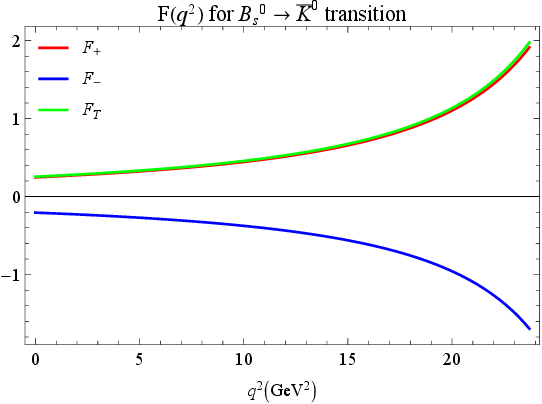}\\
\vspace{0.25cm}
\includegraphics[width=0.45\textwidth]{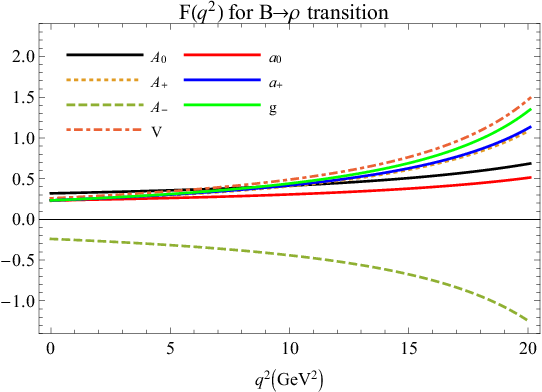}
\hfill\includegraphics[width=0.45\textwidth]{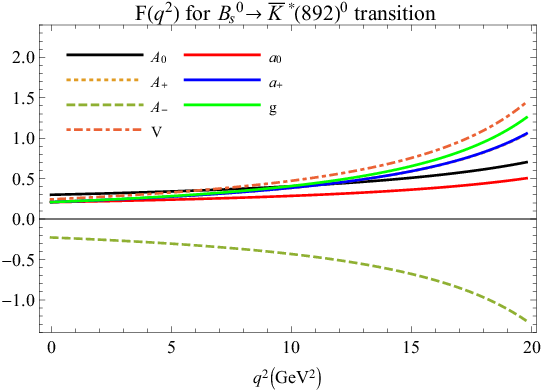}\\
\includegraphics[width=0.45\textwidth]{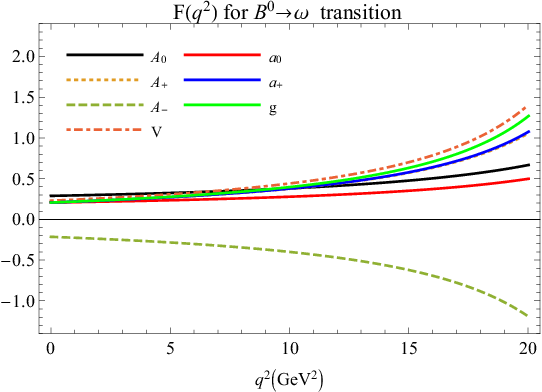}
\vspace{0.25cm}
\caption{Form factors appearing in Eq. (\ref{eq:ff_PP}) - (\ref{eq:ff_PV})}
\label{fig:form_factors}
\end{figure*}

The obvious model parameters include constituent quark masses and meson size parameters that are fixed by fitting with the basic processes such as leptonic decay widths with the experimental data or lattice simulations and the differences are considered to be the absolute uncertainty in the respective parameter.
These uncertainties are observed to be less than 10~\% at maximum recoil which are further transported to the computed form factors and branching fractions.
For present computations, we use the model parameters obtained using the updated least square fit method performed in the Ref. \cite{Ivanov:2015tru,Ganbold:2014pua,Dubnicka:2016nyy}.
After defining the parameters in Tab. \ref{tab:ccqm_parameters}, the form factors appearing in Eqns. (\ref{eq:ff_PP}), (\ref{eq:ff_PV}) are computed using the FORTRAN and  \textit{Mathematica} code.
For detailed information regarding the model and computation techniques used for loop and multidimensional integrals, we suggest the reader to refer to Refs. \cite{Branz:2009cd,Lyubovitskij:2003pn,Ivanov:2019nqd}.
CCQM is a versatile quark model capable for studying the hadronic interaction of multiquark state also and has been recently utilised for computing various decay properties of $D$, $D_s$ and $B_c$ mesons \cite{Ivanov:2019nqd,Soni:2020sgn,Soni:2018adu,Dubnicka:2018gqg,Issadykov:2018myx,Issadykov:2018myx,Soni:2017eug,Soni:2019huk,Soni:2019qjs,Soni:2021rem}, baryons \cite{Gutsche:2019iac,Gutsche:2018msz,Gutsche:2018nks,Gutsche:2018utw} and exotic states \cite{Dubnicka:2020yxy,Dubnickova:2020ljc,Gutsche:2017twh,Goerke:2017svb,Goerke:2016hxf,Gutsche:2016cml}.
In Fig. \ref{fig:form_factors} we provide the computed form factors. The preliminary results on $B \to \rho$ and $B_s^0 \to \bar{K}^*(892)^0$ decay form factors have been calculated in \cite{Issadykov:2019vpm,Issadykov:2019gek}.
\begin{figure*}[htbp]
\centering
\includegraphics[width=0.45\textwidth]{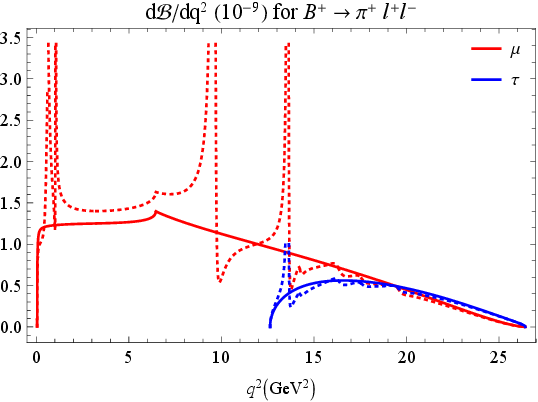}
\hfill\includegraphics[width=0.45\textwidth]{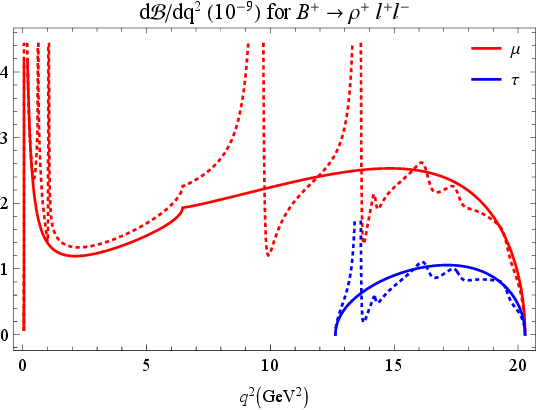}\\
\vspace{0.2cm}
\includegraphics[width=0.45\textwidth]{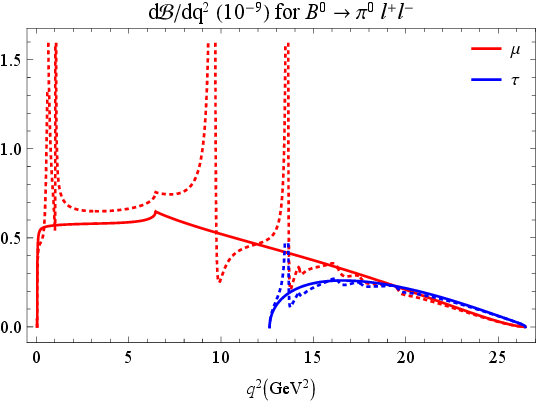}
\hfill\includegraphics[width=0.45\textwidth]{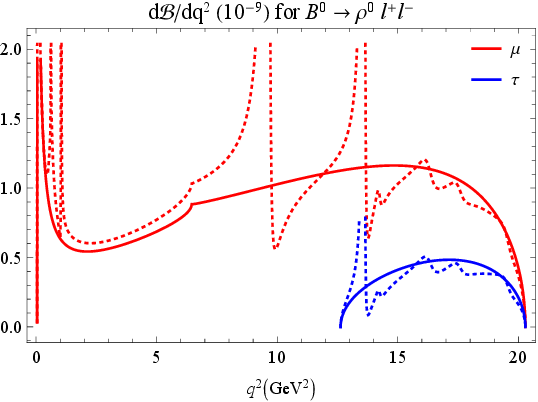}\\
\vspace{0.2cm}
\includegraphics[width=0.45\textwidth]{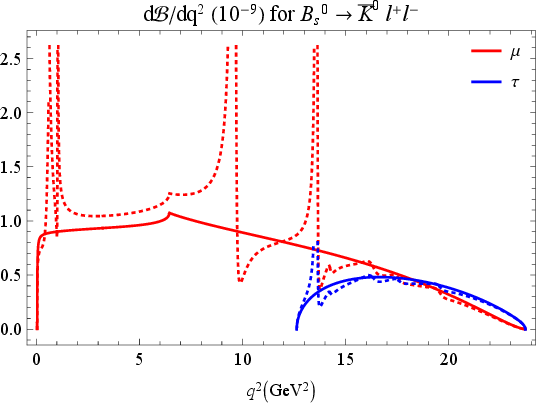}
\hfill\includegraphics[width=0.45\textwidth]{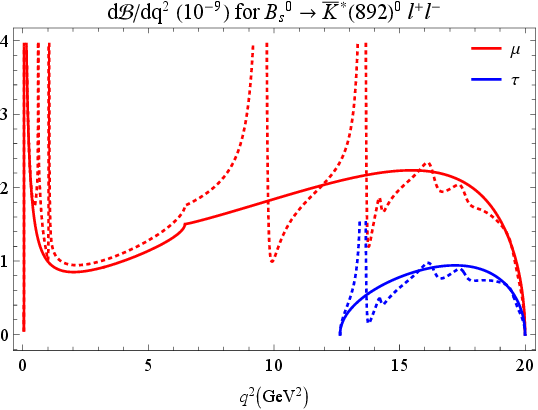}\\
\vspace{0.2cm}
\includegraphics[width=0.45\textwidth]{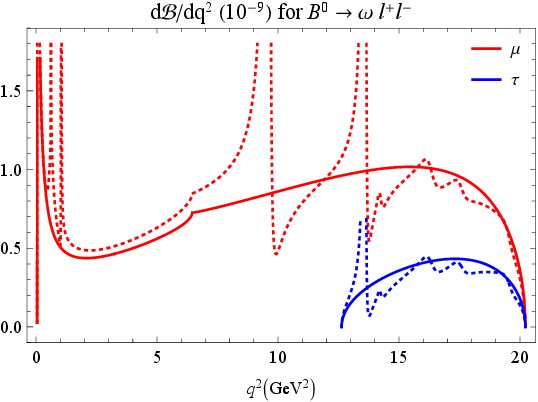}
\caption{Differential branching fractions (solid lines - excluding resonances, dashed lines - including vector resonances).}
\label{fig:defferential_branching}
\end{figure*}

The form factors appearing in Eq. (\ref{eq:ff_PP}) - (\ref{eq:ff_PV}) and plotted in Fig. \ref{fig:form_factors} are also represented in double pole approximation as
\begin{eqnarray}
F(q^2) = \frac{F(0)}{1 - a s + b s^2}, \ \ \ \ s = \frac{q^2}{m_{B_{(s)}}^2}
\label{eq:double_pole}
\end{eqnarray}
and the parameters in the double pole approximation for the different decay channels are given in the Tab. \ref{tab:double_pole_data}.
Note that this double pole parametrization is very precise and relative error for all the form factors with the exact results is less than $1~\%$ for the entire momentum transferred square range.
\begin{table*}[!htbp]
\caption{Form factors and double pole parameters of Eq. \ref{eq:double_pole}.}
\begin{tabular*}{\textwidth}{@{\extracolsep{\fill}}lccclccc@{}}
\hline\hline
$F$ & $F(0)$ & $a$ & $b$ & $F$ & $F(0)$ & $a$ & $b$\\
\hline
$F_+^{B \to \pi}$ & $0.283 \pm 0.019$ & 1.294 & 0.349 & $F_-^{B \to \pi}$ & $-0.238 \pm 0.016$ & 1.323 & 0.374\\
$F_T^{B \to \pi}$ & $0.268 \pm 0.018$ & 1.292 & 0.346 & \\
\hline
$A_+^{B \to \rho}$& $0.227 \pm 0.018$ & 1.355 & 0355 & $A_-^{B \to \rho}$ & $-0.240 \pm 0.019$ & 1.415 & 0.404\\
$A_0^{B \to \rho}$& $0.319 \pm 0.026$ & 0.528 & $-0.295$& $V^{B \to \rho}$ & $0.259 \pm 0.021$ & 1.472 & 0.452\\
$a_0^{B \to \rho}$& $0.233 \pm 0.019$ & 0.575 & $-0.254$ & $a_+^{B \to \rho}$ & $0.233 \pm0.019 $ & 1.362 & 0.360\\
$g^{B^+ \to \rho^+}$	& $0.233 \pm 0.019$ & 1.477 & 0.457\\
\hline
$A_+^{B^0 \to \omega}$	& $0.206 \pm 0.016$	& 1.390 & 0.375 & $A_-^{B^0 \to \omega}$	& $-0.214 \pm 0.017$ & 1.442 & 0.417\\
$A_0^{B^0 \to \omega}$	& $0.288 \pm 0.023$	& 0.557 & $-0.325$ & $V^{B^0 \to \omega}$ & $0.229 \pm 0.023$ & 1.504 & 0.472\\
$a_0^{B^0 \to \omega}$	& $0.206 \pm 0.017$ & 0.618 & $-0.275$ & $a_+^{B^0 \to \omega}$ & $0.206 \pm 0.017$ & 1.401 & 0.384\\
$g^{B^0 \to \omega}$		&	$0.206 \pm 0.017$ & 1.506 & 0.472\\	
\hline
$F_+^{B_s^0\to \bar{K}^0}$ & $0.247 \pm 0.016$ & 1.441 & 0.465 & $F_-^{B_s^0\to \bar{K}^0}$ & $-0.205 \pm 0.013$ & 1.474 & 0.494\\
$F_T^{B_s^0\to \bar{K}^0}$ & $0.256 \pm 0.016$ & 1.429 & 0.451 & \\
\hline
$A_+^{B_s^0 \to \bar{K}^*(892)^0}$& $0.210 \pm 0.015$ & 1.463 & 0.435 &$A_-^{B_s^0 \to \bar{K}^*(892)^0}$ & $-0.228 \pm 0.016$ & 1.539 & 0.504\\
$A_0^{B_s^0 \to \bar{K}^*(892)^0}$& $0.300 \pm 0.021$ & 0.654 & $-0.262$ & $V^{B_s^0 \to \bar{K}^*(892)^0}$ & $0.244 \pm 0.018$ & 1.584 & 0.545\\
$a_0^{B_s^0 \to \bar{K}^*(892)^0}$& $0.210 \pm 0.015$ & 0.706 & $-0.214$ & $a_+^{B_s^0 \to \bar{K}^*(892)^0}$ & $0.210 \pm 0.015$ & 1.470 & 0.441\\
$g^{B_s^0 \to \bar{K}^*(892)^0}$ & $0.210 \pm 0.015$ & 1.597 & 0.559\\
\hline\hline
\end{tabular*}
\label{tab:double_pole_data}
\end{table*}
Using the form factors Tab. \ref{tab:double_pole_data}, model parameters Tab. \ref{tab:ccqm_parameters} and Wilson coefficients Tab. \ref{tab:input}, we compute the branching fractions for rare $b \to d \ell^+ \ell^-$ decays. The width of those decays are computed by integration of the  $q^2$-differential distribution \cite{Faessler:2002ut}
\begin{eqnarray}
\frac{d\Gamma(b \to  d \ell^+ \ell^-)}{dq^2} = \frac{G^2_F}{(2\pi)^3} \left(\frac{\alpha \lambda_d}{2 \pi}\right)^2  \frac{|{\bf p_2}| q^2 \beta_\ell}{12 m_1^2} \mathcal{H}_{\mathrm{tot}}
\label{eq:branching}
\end{eqnarray}
where
\begin{eqnarray}
\mathcal{H}_{\mathrm{tot}}  &=&  \frac12 (\mathcal{H}_U^{11} + \mathcal{H}_U^{22} + \mathcal{H}_L^{11} + \mathcal{H}_L^{22})  +  \delta_{\ell\ell} \left(\frac12 \mathcal{H}_U^{11} - \mathcal{H}_U^{22} + \frac12 \mathcal{H}_L^{11} - \mathcal{H}_L^{22} + \frac32 \mathcal{H}_S^{22} \right).
\label{eq:branching_bilinear}
\end{eqnarray}
In what follows, we use the short notation $m_1=m_{B_{(s)}}$ and $m_2$ is the mass of daughter mesons,  $\beta_\ell=\sqrt{1-4m_\ell^2/q^2}$, $\delta_{\ell\ell} = 2m^2_\ell/q^2$ is the helicity flip suppression factor.
Then  $|{\bf p_2}|=\lambda^{1/2}(m_1^2,m_2^2,q^2)/(2\,m_1)$ is the momentum of
the  daughter meson in the $B_{(s)}$-rest frame with the K\"allen function $\lambda (a, b, c) = a^2 + b^2 + c^2 - 2 (a b + b c + c a)$. Also $\lambda_d = |V_{tb}^* V_{td}|$ is the product of CKM matrix elements.

In the above Eq. (\ref{eq:branching_bilinear}), the bilinear combinations of the helicity structure function for $B_{(s)}^{(0)} \to \pi, \bar{K}^0$ decay channels are defined as \cite{Faessler:2002ut},
\begin{eqnarray}
\mathcal{H}^{ii}_U   &=&  0, \qquad
\mathcal{H}^{ii}_L   = |H^i_{0}|^2, \qquad
\mathcal{H}^{ii}_S   = |H^i_{t0}|^2 .
\end{eqnarray}
The helicity amplitudes in terms of form factors are expressed as,
\begin{eqnarray}
H^i_0 & = & \frac{2 m_1 |{\bf p_2}|}{\sqrt{q^2}} \mathcal{F}_+^i, \nn
H^i_{t0} & = & \frac{1}{\sqrt{q^2}} ((m_1^2 - m_2^2) \mathcal{F}_+^i + q^2 \mathcal{F}_-^i)
\end{eqnarray}
and the form factors $\mathcal{F}_{+-}^i$ for $i = 1, 2$ are related to form factors Eq. (\ref{eq:ff_PP}) as
\begin{eqnarray}
\mathcal{F}_+^1 &=& C_9^{\mathrm{eff}} F_+ + C_7^{\mathrm{eff}} F_T \frac{2 \bar{m_b}}{m_1  +m_2}, \nn
\mathcal{F}_-^1 &=& C_9^{\mathrm{eff}} F_- - C_7^{\mathrm{eff}} F_T \frac{2 \bar{m_b}}{m_1  +m_2} \frac{m_1^2 - m_2^2}{q^2}, \nn
\mathcal{F}_+^2 &=& C_{10} F_+\ \ , \qquad \mathcal{F}_-^2 = C_{10} F_-.
\end{eqnarray}

Similarly, the bilinear combinations of the helicity structure function for $B_{(s)}^{(0)} \to \rho, \omega, \bar{K}^*(892)^0$ decay channels are defined as \cite{Faessler:2002ut},
\begin{eqnarray}
\mathcal{H}^{ii}_U   & = &  |H^i_{+1 +1}|^2 +  |H^i_{-1 -1}|^2, \nn
\mathcal{H}^{ii}_L   & = & |H^i_{00}|^2, \qquad
\mathcal{H}^{ii}_S   = |H^i_{t0}|^2 ,
\end{eqnarray}
where the helicity amplitudes are expressed via the form factors appearing in the matrix element of the $b \to d \ell^+\ell^-$ rare decay as
\begin{eqnarray}
H^i_{t0} &=&
\frac{1}{m_1+m_2}\frac{m_1\,|{\bf p_2}|}{m_2\sqrt{q^2}}
         \left(Pq\,(-A^i_0+A^i_+)+q^2 A^i_-\right), \nn
H^i_{\pm1\pm1} &=&
\frac{1}{m_1+m_2}\left(-Pq\, A^i_0\pm 2\,m_1\,|{\bf p_2}|\, V^i \right), \nn
H^i_{00} &=&
\frac{1}{m_1+m_2}\frac{1}{2\,m_2\sqrt{q^2}}  
\left(-Pq\,(m_1^2 - m_2^2 - q^2)\, A^i_0 + 4\,m_1^2\,|{\bf p_2}|^2\, A^i_+\right).
\label{eq:hel_V}
\end{eqnarray}
The form factors $A^i$ and $V^i$ $(i=1,2)$ are related to the transition form factors for the decay $b \to d$ Eq. (\ref{eq:ff_PV}) in the following manner
\begin{eqnarray}
V^{(1)} &=&   C_9^{\rm eff}\,V  + C_7^{\rm eff}\,g \,\frac{2\bar m_b(m_1+m_2)}{q^2}\,,
\nn
A_0^{(1)} &=& C_9^{\rm eff}\,A_0
+ C_7^{\rm eff}\,a_0\,\frac{2\bar m_b(m_1+m_2)}{q^2}\,,
\nn
A_+^{(1)} &=& C_9^{\rm eff}\,A_+
+ C_7^{\rm eff}\,a_+\,\frac{2\bar m_b(m_1+m_2)}{q^2}\,,
\nn
A_-^{(1)} &=& C_9^{\rm eff}\,A_-
+ C_7^{\rm eff}\,(a_0-a_+)\,\frac{2\bar m_b(m_1+m_2)}{q^2}\,\frac{Pq}{q^2}\,,
\nn[1.5ex]
V^{(2)}   &=& C_{10}\,V, \qquad A_0^{(2)} = C_{10}\,A_0,\qquad
A_\pm^{(2)} = C_{10}\,A_\pm.
\label{eq:ff-relations}
\end{eqnarray}
Having defined the helicity structure functions, we plot the differential branching fractions using Eq. (\ref{eq:branching}) in Fig. \ref{fig:defferential_branching}. The corresponding rare branching fractions are computed by numerical integration of Fig. \ref{fig:defferential_branching} and tabulated in Tab. \ref{tab:branching_B} and \ref{tab:branching_Bs}.

We also compute the branching fractions corresponding to $b \to d \nu \bar{\nu}$ decays. The differential branching fractions are expressed as \cite{Faessler:2002ut}
\begin{eqnarray}
\frac{d\Gamma(b \to d \nu\bar\nu)}{dq^2}
&=& \frac{G_F^2}{(2\pi)^3} \Big(\frac{\alpha\lambda_d}{2\pi}\Big)^2
\left[\frac{D_\nu(x_t)}{\sin^2\theta_W}\right]^2
\frac{|{\bf p_2}|\, q^2}{4m_1^2} \cdot (H_U+H_L),
\label{eq:rare_nu}
\end{eqnarray}
where $x_t=\bar m_t^2/m_W^2$ and the function $D_\nu$ with $\alpha_s$ correction is given by \cite{Buchalla:1998ba}
\begin{eqnarray}
D_\nu(x) = D_0(x) + \frac{\alpha_s}{4\pi} D_1(x)
\end{eqnarray}
with
\begin{eqnarray}
D_0(x) = \frac{x}{8}\left(\frac{2+x}{x-1}+\frac{3x-6}{(x-1)^2}\,\ln x\right)
\end{eqnarray}
and
\begin{eqnarray}
D_1(x) &=& - \frac{29 x - x^2 - 4 x^3}{3 (1-x)^2} - \frac{x + 9 x^2 - x^3 - x^4}{(1-x)^3}~ \ln x \nn & + & \frac{8 x + 4 x^2 + x^3 - x^4}{2 (1-x)^3}~\ln^2 x -  \frac{4 x - x^3}{(1-x)^2} \int_1^x dt \frac{\ln t}{1-t} \nn & + &  8 x \frac{\partial D_0(x)}{\partial x} ~\ln \left(\frac{\mu_b^2}{m_W^2}\right).
\end{eqnarray}
The relevant bilinear helicity combinations for the channels $B_{(s)}^{(0)} \to \pi, \bar{K}^0$ can be written as
\begin{eqnarray}
\mathcal{H}_L = |H_0|^2,  \qquad \mathcal{H}_U = 0
\end{eqnarray}
with
\begin{eqnarray}
H_0 = \frac{2 m_1 |\textbf{p}_2|}{\sqrt{q^2}} F_+.
\label{eq:bilinear_2_PP}
\end{eqnarray}
Similarly, the bilinear helicity combinations for the channels $B_{(s)}^{(0)} \to \rho, \omega, \bar{K}^*(892)^0$ can be expressed as
\begin{eqnarray}
\mathcal{H}_U   &=&  |H_{+1 +1}|^2 +  |H_{-1 -1}|^2, \qquad
\mathcal{H}_L   = |H_{00}|^2,
\label{eq:bilenear_HuHl}
\end{eqnarray}
with
\begin{eqnarray}
H_{\pm1\pm1} &=&
\frac{1}{m_1+m_2}\left(-Pq\, A_0\pm 2\,m_1\,|{\bf p_2}|\, V \right),
\nn
H_{00} &=&
\frac{1}{m_1+m_2}\frac{1}{2\,m_2\sqrt{q^2}} 
\left(-Pq\,(m_1^2 - m_2^2 - q^2)\, A_0 + 4\,m_1^2\,|{\bf p_2}|^2\, A_+\right).
\label{eq:bilinear_2_PV}
\end{eqnarray}
For computation of branching fractions of Eq. (\ref{eq:rare_nu}), the form factors ($F_+$, $A_{0,+,-}$ and $V$) appearing in Eq. (\ref{eq:bilinear_2_PP}) and (\ref{eq:bilinear_2_PV}) are taken from Tab. \ref{tab:double_pole_data}.

Finally, we compute the radiative decay width $B_{(s)}^{(0)} \to (\rho, \omega, \bar{K}^*(892)^0) \gamma $ using the relation
\begin{eqnarray}
\Gamma(B_{(s)}^{(0)} \to (\rho, \omega, \bar{K}^*(892)^0) \gamma)
&=& \frac{G_F^2\alpha |V_{tb} V_{td}^*|^2}{32\pi^4}
\bar m_b^2 m_1^3 \left(1-\frac{m_2^2}{m_1^2}\right)^3\,|C^{\rm eff}_7|^2\, g^2(0).
\end{eqnarray}
where, $\alpha$ is the electromagnetic coupling constant (Tab. \ref{tab:input}).

\section{Results and Discussion}
\label{sec:result}
Having determined the model parameters in Tab. \ref{tab:ccqm_parameters}, the transition form factors Eq. (\ref{eq:ff_PP}) and (\ref{eq:ff_PV}) for the rare $B_{(s)}$ decays are computed in the entire dynamical range of momentum transfer and plotted in Fig. \ref{fig:form_factors}.
We also compare our form factors with other theoretical approaches. In order to compare with other theoretical approaches, we relate our form factors Eq. (\ref{eq:ff_PP}) and (\ref{eq:ff_PV}) to those with Bauer-Stech-Wirbel (BSW) form factors \cite{Wirbel:1985ji}. We denote them by the superscript $^c$ to distinguish from our form factors. The relations read,
\begin{eqnarray}
F_0^c & = & F_+ + \frac{q^2}{m_1^2 - m_2^2} F_-,\\
A_0 &=& \frac{m_1 + m_2}{m_1 - m_2}\,A_1^c\,, \qquad
A_+ = A_2^c\,,
\nn
A_- &=&  \frac{2m_2(m_1+m_2)}{q^2}\,(A_3^c - A_0^c)\,, \qquad
V = V^c\,,
\nn[1.2ex]
a_0 &=& T_2^c\,, \qquad g = T_1^c\,, \qquad
a_+  =  T_2^c + \frac{q^2}{m_1^2-m_2^2}\,T_3^c\,.
\label{eq:new-ff}
\end{eqnarray}
Additionally, we also note that the form factors Eq. (\ref{eq:new-ff}) satisfy the constraints
\begin{eqnarray}
 A_0^c(0) &=& A_3^c(0) \nn
2m_2A_3^c(q^2) &=& (m_1+m_2) A_1^c(q^2) -(m_1-m_2) A_2^c(q^2)\,.
\end{eqnarray}
Since $a_0(0)=a_+(0)=g(0)$, we present the form factors
$A_0^c(0)=(m_1-m_2)[A_0(0)-A_{+}(0)]/(2m_2)$,
$A_1^c(0)=A_0(0)(m_1-m_2)/(m_1+m_2)$,
$A_2^c(0)=A_+(0)$,
$T_1^c(0)=g(0)$ and
$T_3^c(0)=\lim_{\,q^2 \to 0}(m_1^2-m_2^2)(a_{+}-a_0)/q^2$
obtained in our model and compare them with those from other approaches.
Note that for comparing with the other approaches, we omit the superscript for simplification.

\begin{table*}[!htbp]
\caption{$B \to \pi$ and $B_s^0 \to \bar{K}^0$ form factors at maximum recoil}
\begin{tabular*}{\textwidth}{@{\extracolsep{\fill}}c|cc|cc@{}}
\hline\hline
Theory & \multicolumn{2}{c|}{$B \to \pi$} & \multicolumn{2}{c}{$B_s^0 \to \bar{K}^0$}\\
& $f_{+,0} (0)$ & $f_T(0)$ & $f_{+,0} (0)$ & $f_T(0)$ \\
\hline
Present & $0.283 \pm 0.019$ & $0.268 \pm 0.018$ & $ 0.247 \pm 0.015$  & $0.256 \pm 0.016$ \\
LCSR \cite{Lu:2018cfc}& 0.280 & 0.260 & 0.364 & 0.363\\
LCSR \cite{Wu:2006rd} & $0.285^{+0.016}_{-0.015}$ & $0.267^{+0.015}_{-0.014}$ & $0.296 \pm 0.018$ & $0.288^{+0.018}_{-0.017}$\\
LCSR \cite{Khodjamirian:2017fxg} & $0.301 \pm 0.023$ & $0.273 \pm 0.021$ & $0.336 \pm 0.023$ & $0.320 \pm 0.019$\\
LCSR \cite{Gubernari:2018wyi} & $0.21 \pm 0.07$ & $0.19 \pm 0.06$ & -- & --\\
SUSY \cite{Wang:2007sp} & 0.258 & 0.253 & -- & --\\
pQCD \cite{Wang:2012ab} & $0.26^{+0.04}_{-0.03} \pm 0.03 \pm 0.02$ & $0.26^{+0.04}_{-0.03} \pm 0.03 \pm 0.02$ & $0.26^{+0.04}_{-0.03} \pm 0.03 \pm 0.02$ & $0.28 \pm 0.04 \pm 0.03 \pm 0.02$\\
pQCD \cite{Jin:2020jtu} & -- & -- & 0.22  & 0.22\\
SCET \cite{Lu:2007sg} & 0.247 & 0.253 & 0.297 & 0.325 \\
RQM \cite{Faustov:2014zva,Faustov:2013ima} & $0.217 \pm 0.011$ & $0.240 \pm 0.012$ & 0.284 & 0.236\\
CQM \cite{Melikhov:2000yu} & 0.29 & 0.28 & 0.31 & 0.31\\
LFQM \cite{Verma:2011yw} & 0.25 & -- & 0.23 & --\\
\hline\hline
\end{tabular*}
\label{tab:ff_comparison_BP}
\end{table*}

\begin{table*}[!htbp]
\caption{$B \to \rho$ form factors at maximum recoil}
\begin{tabular*}{\textwidth}{@{\extracolsep{\fill}}lcccccc@{}}
\hline\hline
&  $V(0)$ & $A_0(0)$ & $A_1(0)$ & $A_2(0)$ & $T_{1,2}(0)$ & $T_3(0)$\\
\hline
Present & $0.259 \pm 0.021$ & $0.266 \pm 0.013$ & $0.238 \pm 0.019$ & $0.227 \pm 0.018$ & $0.233 \pm 0.019$ & $0.179 \pm 0.014$\\
LCSR \cite{Wu:2006rd} & $0.289 \pm 0.016$ & -- & $0.232^{+0.013}_{-0.014}$ & $0.187^{+0.011}_{-0.012}$ & $0.256 \pm 0.015$ & $0.175 \pm 0.010$\\
LCSR \cite{Ball:2004rg} & 0.323 & 0.303 & 0.242 & 0.221 & 0.267 & 0.176\\
LCSR \cite{Straub:2015ica} & $0.327 \pm 0.031$& $0.356 \pm 0.042$ & $0.262 \pm 0.026$ & $0.297 \pm 0.035$& $0.272 \pm 0.026$ & $0.747 \pm 0.076$\\
LCSR \cite{Gubernari:2018wyi} & $0.27 \pm 0.14$ & -- & -- & $0.19 \pm 0.11$ & $0.24 \pm 0.12$ & --\\
pQCD \cite{Li:2009tx} & $0.21^{+0.05+0.03}_{-0.04-0.02}$  & $0.25^{+0.06+0.04}_{-0.05-0.03}$ & $0.16^{+0.04+0.02}_{-0.03-0.02}$ & $0.13^{+0.03+0.02}_{-0.03-0.01}$ & $0.19^{+0.04+0.03}_{-0.04-0.02}$ & $0.17^{+0.04+0.02}_{-0.03-0.02}$  \\
SCET \cite{Lu:2007sg} & 0.298 & 0.260 & 0.227 & 0.215 & 0.260 & 0.184\\
RQM \cite{Faustov:2014zva} & $0.295 \pm 0.015$ & $0.231 \pm 0.012$ & $0.269 \pm 0.014$ & $0.282 \pm 0.014$ & $0.290 \pm 0.015$ & $0.124 \pm 0.007$\\
CQM \cite{Melikhov:2000yu} & 0.31 & 0.30 & 0.26 & 0.24 & 0.27 & 0.19\\
LFQM \cite{Chang:2019mmh} & $0.35^{+0.01+0.06}_{-0.01-0.05}$ &$0.30^{+0.01+0.05}_{-0.01-0.05}$ & $0.27^{+0.01+0.05}_{-0.01-0.04}$ &$0.25^{+0.01+0.04}_{-0.01-0.04}$& -- & --\\
\hline\hline
\end{tabular*}
\label{tab:ff_comparison_Brho}
\end{table*}

\begin{table*}[!htbp]
\caption{$B^0 \to \omega$ form factors at maximum recoil}
\begin{tabular*}{\textwidth}{@{\extracolsep{\fill}}lcccccc@{}}
\hline\hline
&  $V(0)$ & $A_0(0)$ & $A_1(0)$ & $A_2(0)$ & $T_{1,2}(0)$ & $T_3(0)$\\
\hline
Present & $0.229 \pm 0.023$ & $0.236 \pm 0.011$ & $0.214 \pm 0.017$ & $0.206 \pm 0.016$ & $0.206 \pm 0.017$ & $0.158 \pm 0.013$\\
LCSR \cite{Wu:2006rd} & $0.268^{+0.014}_{-0.015}$ & -- & $0.214^{+0.013}_{-0.012}$ & $0.170^{+0.010}_{-0.011}$ & $0.237^{+0.013}_{-0.014}$ & $0.160 \pm 0.009$\\
LCSR \cite{Ball:2004rg} & 0.293 & 0.281 & 0.219 & 0.198 & 0.242 & 0.155\\
LCSR \cite{Straub:2015ica} & $0.304 \pm 0.038$ & $0.328 \pm 0.048$ & $0.243 \pm 0.031$ & $0.270 \pm 0.040$ & $0.251 \pm 0.031$ & $0.683 \pm 0.090$\\
pQCD \cite{Li:2009tx} & $0.19^{+0.04+0.03}_{-0.04-0.02}$ & $0.23^{+0.05+0.03}_{-0.04-0.02}$ & $0.15^{+0.03+0.02}_{-0.03-0.01}$ & $0.12^{+0.03+0.02}_{-0.02-0.01}$ & $0.18^{+0.04+0.02}_{-0.04-0.02}$ & $0.15^{+0.04+0.02}_{-0.03-0.02}$\\
SCET \cite{Lu:2007sg} & 0.275 & 0.240 & 0.209 & 0.198 & 0.239 & 0.168\\
LFQM \cite{Verma:2011yw} & $0.27$ & $0.28 \pm 0.01$ & 0.23 & 0.21 & -- & --\\
\hline\hline
\end{tabular*}
\label{tab:ff_comparison_Bomega}
\end{table*}

\begin{table*}[!htbp]
\caption{$B_s^0 \to \bar{K}^*(892)^0$ form factors at maximum recoil}
\begin{tabular*}{\textwidth}{@{\extracolsep{\fill}}lccccccc@{}}
\hline\hline
&  $V(0)$ & $A_0(0)$ & $A_1(0)$ & $A_2(0)$ & $T_{1,2}(0)$ & $T_3(0)$\\
\hline
Present & $0.244 \pm 0.018$ & $0.225 \pm 0.090$ & $0.214 \pm 0.015$ & $0.210 \pm 0.015$ & $0.210 \pm 0.015$ & $0.156 \pm 0.011$\\
LCSR \cite{Wu:2006rd} & $0.285 \pm 0.013$ & -- & $0.227^{+0.010}_{-0.012}$ & $0.183^{+0.008}_{-0.010}$ & $0.251 \pm 0.012$ & $0.169 \pm 0.008$\\
LCSR \cite{Ball:2004rg} & 0.31 & 0.36 & 0.23 & 0.18 & 0.26 & 0.14\\
LCSR \cite{Straub:2015ica} & $0.296 \pm 0.030$ & $0.314 \pm 0.048$ & $0.230 \pm 0.025$ & $0.229 \pm 0.035$ & $0.239 \pm 0.024$ & $0.597 \pm 0.076$\\
pQCD \cite{Li:2009tx} & $0.20^{+0.04+0.03}_{-0.04-0.02}$ & $0.24^{+0.05+0.04}_{-0.04-0.02}$ & $0.15^{+0.03+0.02}_{-0.03-0.01}$ & $0.11^{+0.02+0.01}_{-0.02-0.01}$ & $0.18^{+0.04+0.02}_{-0.03-0.0}$ & $0.16^{+0.03+0.02}_{-0.03-0.02}$\\
pQCD \cite{Jin:2020jtu} & 0.24 & 0.21 & 0.19 & 0.19 & 0.21 & 0.16\\
SCET \cite{Lu:2007sg} & 0.323 & 0.279 & 0.228 & 0.204 & 0.271 & 0.165\\
RQM \cite{Faustov:2013ima} & 0.291 & 0.289 & 0.287 & 0.286 & 0.238 & 0.122\\
CQM \cite{Melikhov:2000yu} & 0.38 & 0.37 & 0.29 & 0.26 & 0.32 & 0.23\\
LFQM \cite{Chang:2019mmh} & $0.28^{+0.02+0.07}_{-0.02-0.06}$ &$0.22^{+0.01+0.06}_{-0.01-0.05}$ & $0.20^{+0.01+0.05}_{-0.01-0.05}$ &$0.19^{+0.01+0.05}_{-0.01-0.04}$& -- & --\\
\hline\hline
\end{tabular*}
\label{tab:ff_comparison_BsKv}
\end{table*}

\begin{table*}
\caption{Comparison of form factors at higher $q^2$ values with LQCD (data from Table VI of Ref. \cite{Flynn:2015mha}) for $B_s^0 \to \bar{K}^0$ and $B \to \pi$ channels}
\begin{tabular*}{\textwidth}{@{\extracolsep{\fill}}cc|cc|cc@{}}
\hline\hline
Channel & $q^2$ & \multicolumn{2}{c|}{$F_+(q^2)$} & \multicolumn{2}{c}{$F_0(q^2)$}\\
\cline{3-6}
&GeV$^2$ & Present & LQCD & Present & LQCD \\
\hline
 									& 17.6 & $0.84 \pm 0.05$ & $0.99 \pm 0.06$ 	& $0.40 \pm 0.03$	& $0.48 \pm 0.03$\\
$B_s^0 \to \bar{K}^0$ 	& 20.8 & $1.22 \pm 0.05$ & $1.64 \pm 0.09$ 	& $0.45 \pm 0.03$	& $0.63 \pm 0.04$\\
 									& 23.4 & $1.81 \pm 0.11$ & $2.77 \pm 0.15$	& $0.50 \pm 0.03$ & $0.81 \pm 0.05$\\
\hline
 									& 19.0 & $1.01 \pm 0.07$ & $1.21 \pm 0.13$ 		& $0.41 \pm 0.03$ & $0.46 \pm 0.06$\\
$B^+ \to \pi^+$			 	& 22.6 & $1.57 \pm 0.010$ & $2.27 \pm 0.19$ 	& $0.45 \pm 0.03$ & $0.68 \pm 0.06$\\
 									& 25.1 & $2.40 \pm 0.16$ & $4.11 \pm 0.59$ 		& $0.48 \pm 0.03$ & $0.92 \pm 0.07$\\
\hline\hline
\end{tabular*}
\label{tab:PP_FF_LQCD}
\end{table*}

\begin{table}
\caption{Comparison of form factors at higher $q^2$ (in GeV$^2$) values with LQCD \cite{Horgan:2013hoa} for $B_s^0 \to \bar{K}^*(892)^0$}
\begin{tabular*}{\textwidth}{@{\extracolsep{\fill}}ccccc@{}}
\hline\hline
$F(q^2)$ & & $q^2 = 12$ & $q^2 = 16$ & $q^2 = q^2_{\mathrm{max}} $\\
\hline
$V(q^2)$ & Present & $0.56 \pm 0.04$ & $0.85 \pm 0.06$ & $1.50 \pm 0.10$ \\
			&	LQCD & 0.56 (9) & 1.02 (8) & 1.99 (13)\\
$A_0(q^2)$ & Present & $0.52 \pm 0.06$ & $0.79 \pm 0.09$ & $1.41 \pm 0.13$ \\
			&	LQCD  & 0.84 (9) & 1.33 (8) & 2.38 (16)\\
$A_1(q^2)$ & Present & $0.31 \pm 0.02$ & $0.38 \pm 0.03$ & $0.51 \pm 0.03$ \\
			&	LQCD  & 0.37 (3) & 0.45 (3) & 0.58 (3)\\
$A_2(q^2)$ & Present & $0.45 \pm 0.03$ & $0.65 \pm 0.04$ & $1.08 \pm 0.07$  \\
			&	LQCD  & 0.46 (3) & 0.60 (5) & 0.85 (12)\\
$T_1(q^2)$ & Present & $0.48 \pm 0.03$ & $0.73 \pm 0.05$ & $1.30 \pm 0.09$  \\
			&	LQCD & 0.61 (4) & 0.90 (6) & 1.48 (10)\\
$T_2(q^2)$ & Present & $0.31 \pm 0.02$ & $0.39 \pm 0.03$ & $0.52 \pm 0.04$  \\
			&	LQCD & 0.39 (3) & 0.47 (3) & 0.60 (3)\\
$T_3(q^2)$ & Present & $0.32 \pm 0.04$ & $0.47 \pm 0.05$ & $0.80 \pm 0.06$  \\
			&	LQCD  & 0.43 (4) & 0.67 (5) & 1.10 (7)\\
\hline\hline
\end{tabular*}
\label{tab:PV_FF_LQCD}
\end{table}

\begin{figure*}[!htbp]
\includegraphics[width=0.45\textwidth]{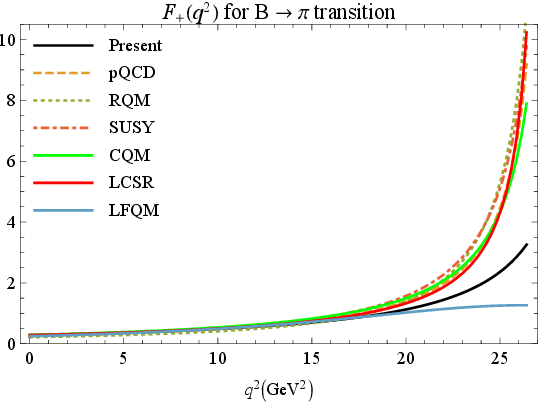}
\hfill\includegraphics[width=0.45\textwidth]{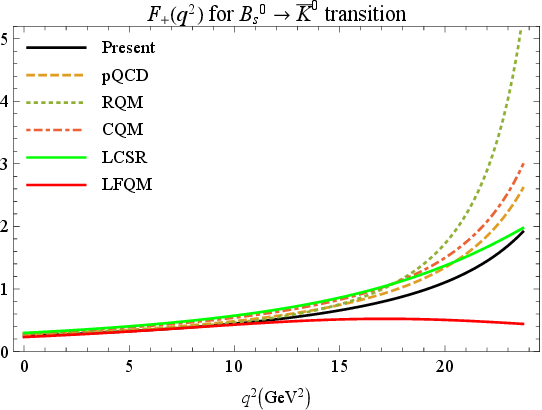}\\
\vspace{0.25cm}
\includegraphics[width=0.45\textwidth]{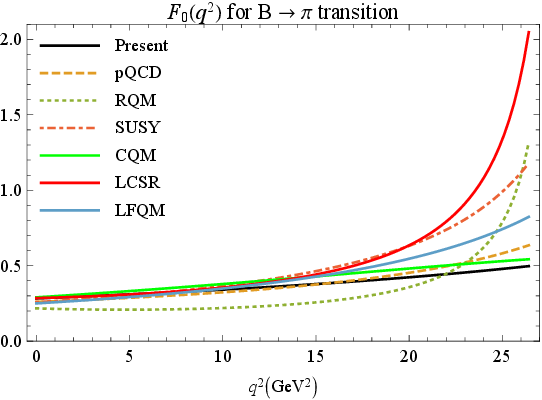}
\hfill\includegraphics[width=0.45\textwidth]{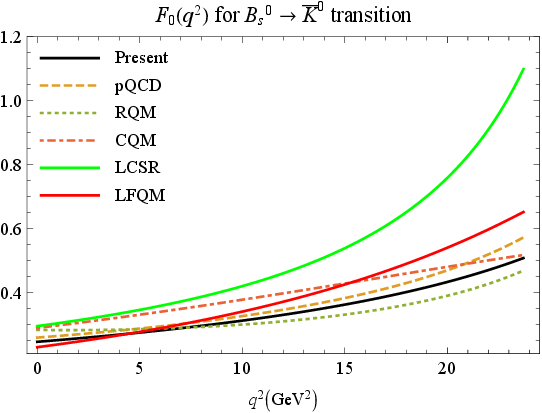}\\
\vspace{0.25cm}
\includegraphics[width=0.45\textwidth]{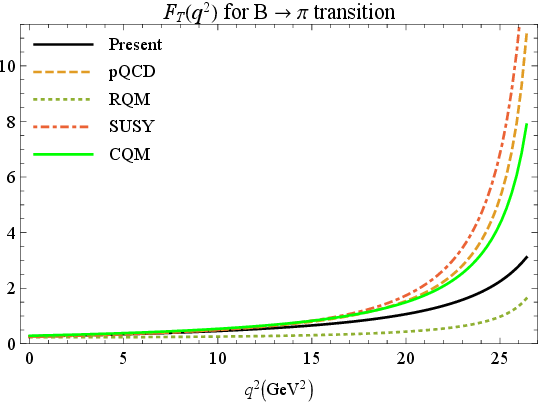}
\hfill\includegraphics[width=0.45\textwidth]{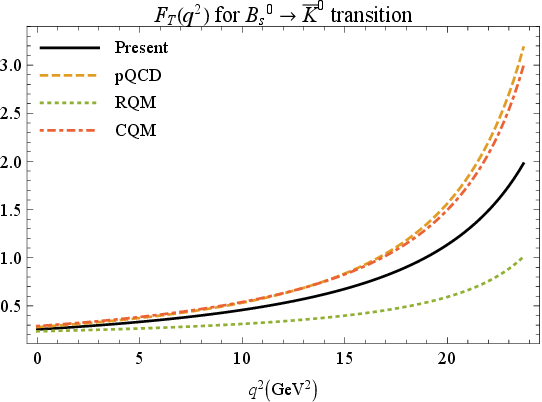}\\
\caption{Form factors comparision for $B^+ \to \pi^+$ (left) and $B_s^0 \to \bar{K}^0$ (right) transition in our model and with pQCD \cite{Wang:2012ab}, RQM \cite{Faustov:2014zva,Faustov:2013ima}, SUSY \cite{Wang:2007sp}, CQM \cite{Melikhov:2000yu}, LCSR \cite{Wu:2006rd}, LFQM \cite{Verma:2011yw}.}
\label{fig:B_P}
\end{figure*}
\begin{figure*}[!htbp]
\includegraphics[width=0.32\textwidth]{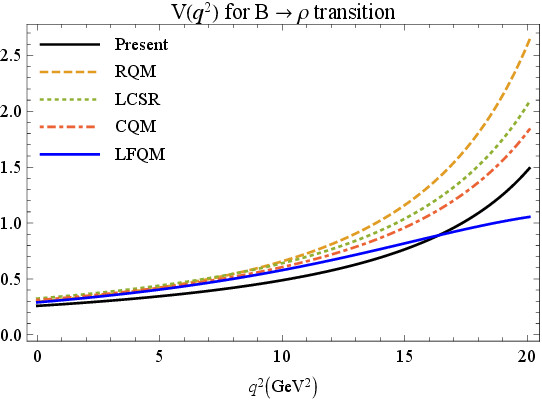}
\hfill\includegraphics[width=0.32\textwidth]{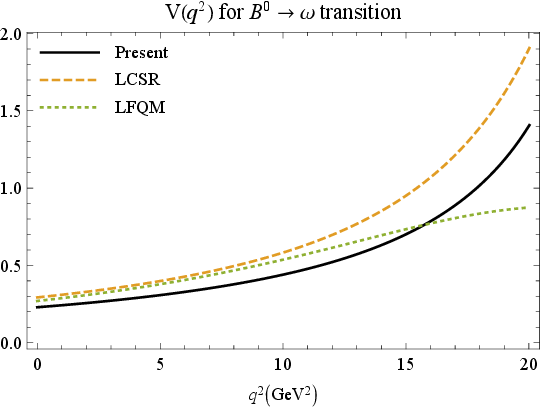}
\hfill\includegraphics[width=0.32\textwidth]{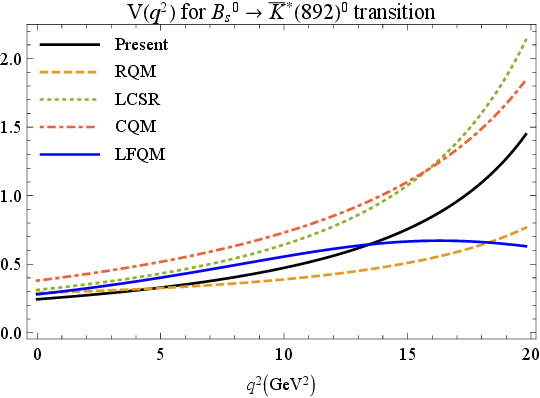}\\
\vspace{0.25cm}
\includegraphics[width=0.32\textwidth]{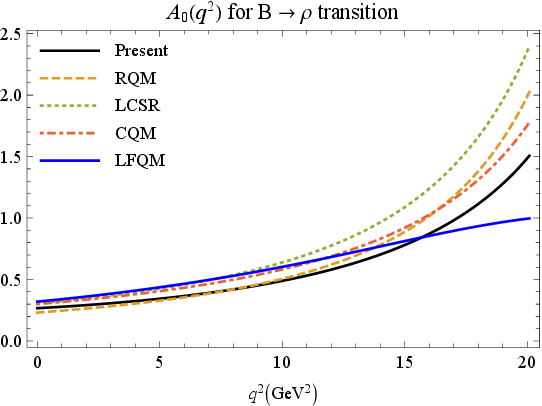}
\hfill\includegraphics[width=0.32\textwidth]{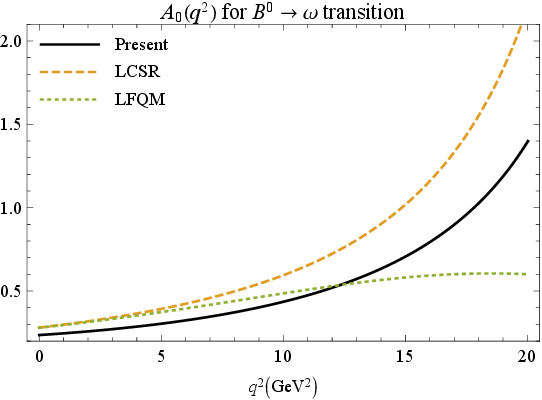}
\hfill\includegraphics[width=0.32\textwidth]{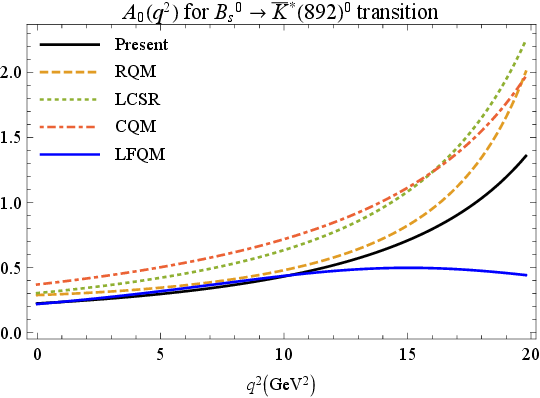}\\
\vspace{0.25cm}
\includegraphics[width=0.32\textwidth]{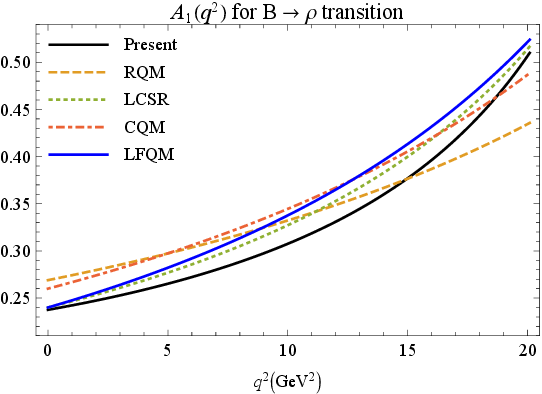}
\hfill\includegraphics[width=0.32\textwidth]{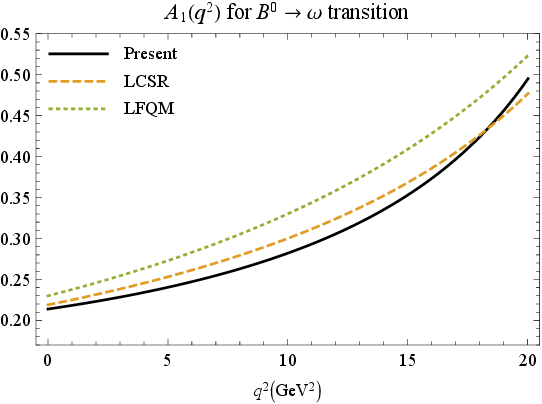}
\hfill\includegraphics[width=0.32\textwidth]{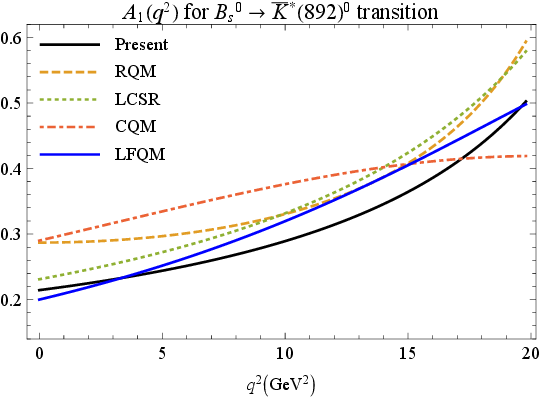}\\
\vspace{0.25cm}
\includegraphics[width=0.32\textwidth]{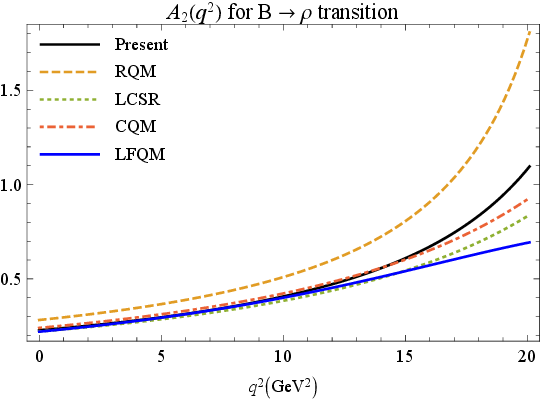}
\hfill\includegraphics[width=0.32\textwidth]{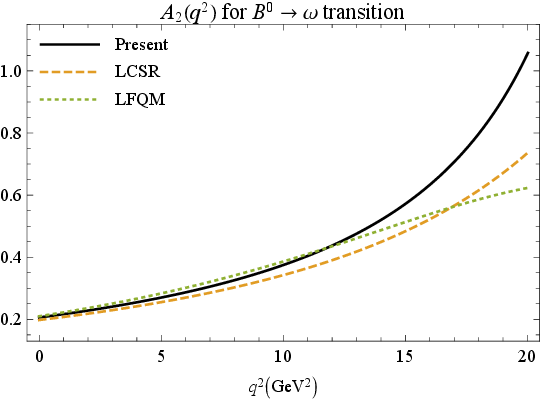}
\hfill\includegraphics[width=0.32\textwidth]{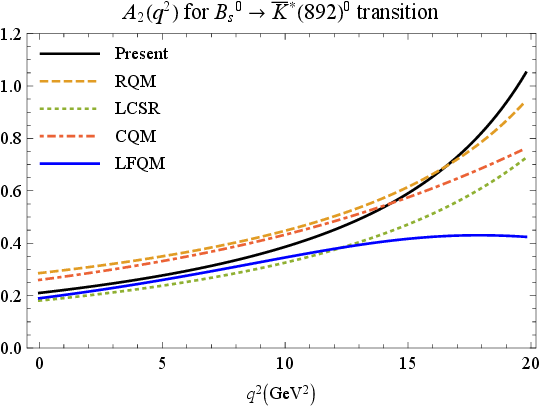}\\
\caption{Form factors comparision for $B \to \rho$ (left), $B^0 \to \omega$ (middle) and $B_s^0 \to \bar{K}^*(892)^0$ (right) transition in our model and with RQM \cite{Faustov:2014zva,Faustov:2013ima}, LCSR \cite{Ball:2004rg}, CQM \cite{Melikhov:2000yu} and LFQM \cite{Chang:2019mmh,Verma:2011yw}.}
\label{fig:B_V}
\end{figure*}
\begin{figure*}[!htbp]
\includegraphics[width=0.32\textwidth]{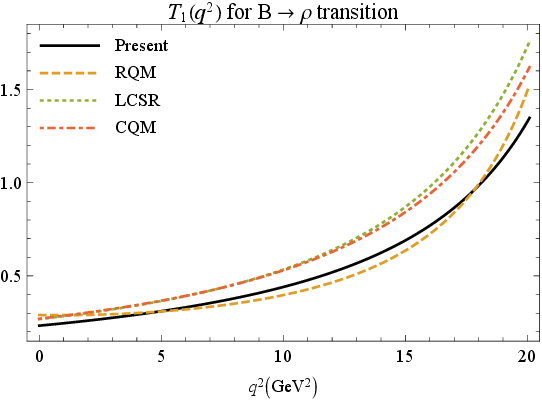}
\hfill\includegraphics[width=0.32\textwidth]{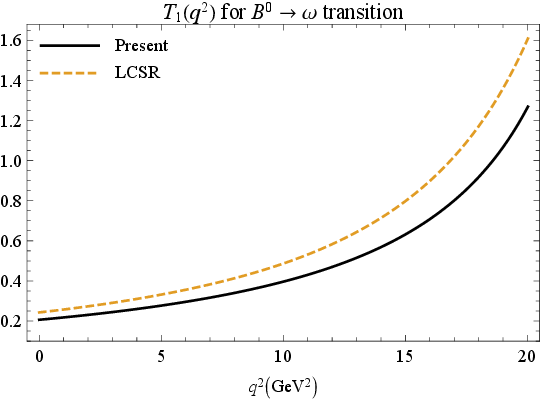}
\hfill\includegraphics[width=0.32\textwidth]{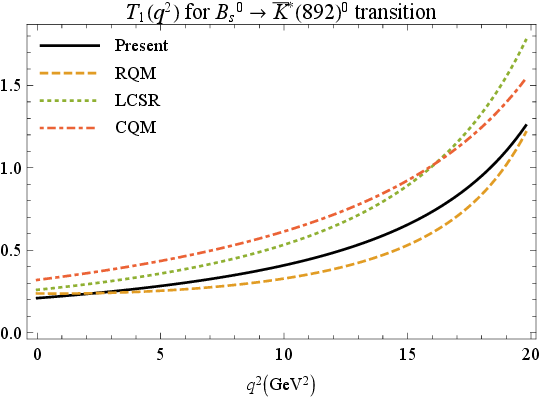}\\
\vspace{0.25cm}
\includegraphics[width=0.32\textwidth]{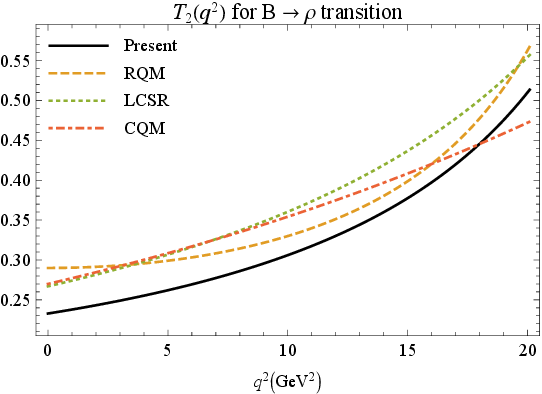}
\hfill\includegraphics[width=0.32\textwidth]{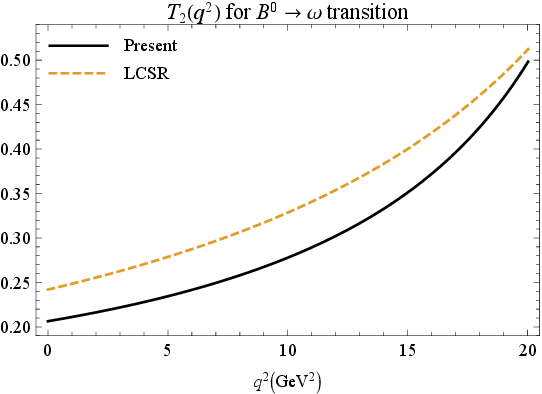}
\hfill\includegraphics[width=0.32\textwidth]{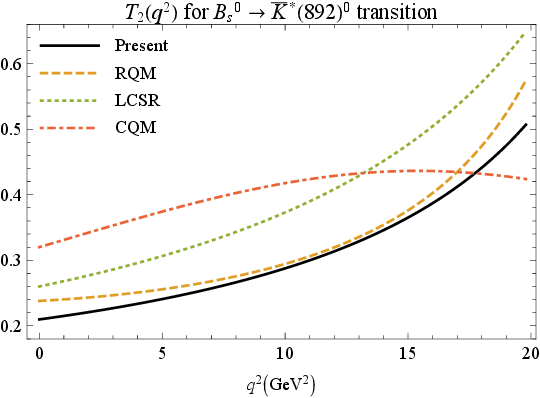}\\
\vspace{0.25cm}
\includegraphics[width=0.32\textwidth]{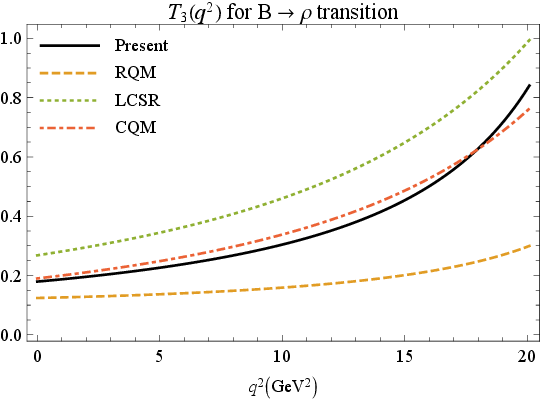}
\hfill\includegraphics[width=0.32\textwidth]{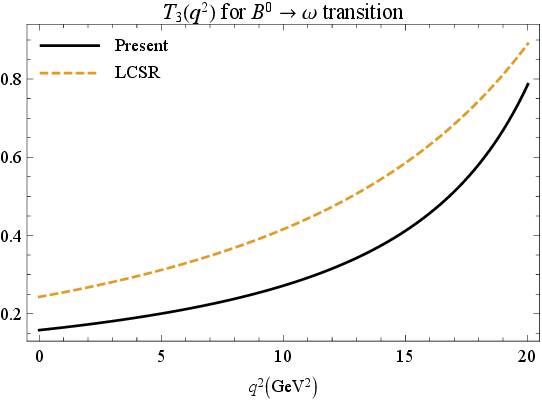}
\hfill\includegraphics[width=0.32\textwidth]{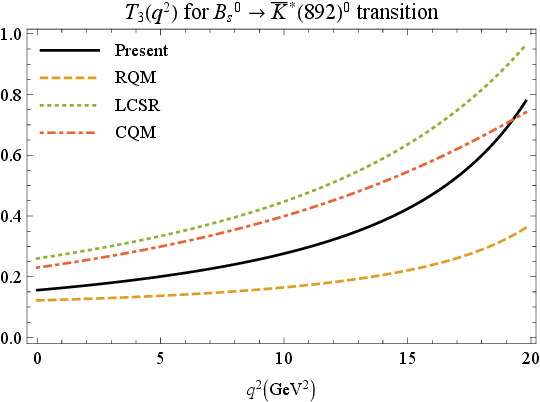}\\
\caption{Form factors comparision for $B \to \rho$ (left), $B^0 \to \omega$ (middle) and $B_s^0 \to \bar{K}^*(892)^0$ (right) transition in our model and with RQM \cite{Faustov:2014zva,Faustov:2013ima}, LCSR \cite{Ball:2004rg} and CQM \cite{Melikhov:2000yu}.}
\label{fig:B_V_Ts}
\end{figure*}
\begin{figure*}[!htbp]
\includegraphics[width=0.45\textwidth]{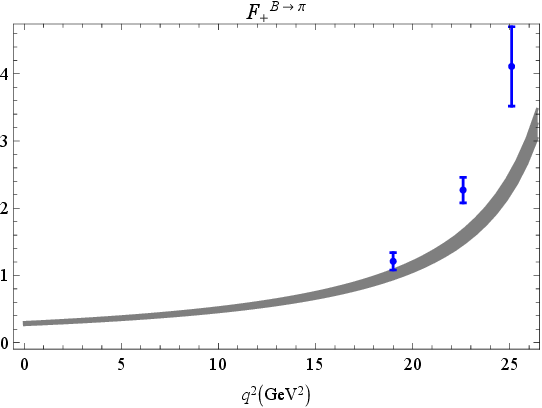}
\hfill\includegraphics[width=0.45\textwidth]{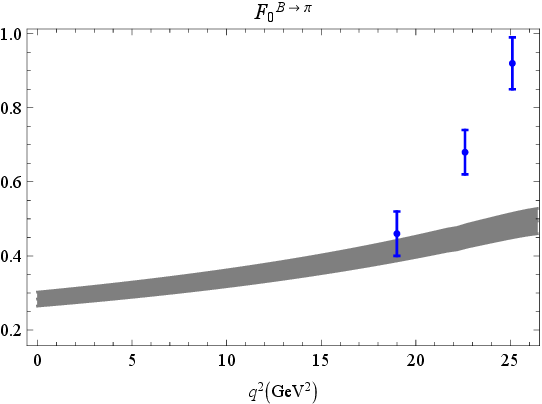}\\
\vspace{0.25cm}
\includegraphics[width=0.45\textwidth]{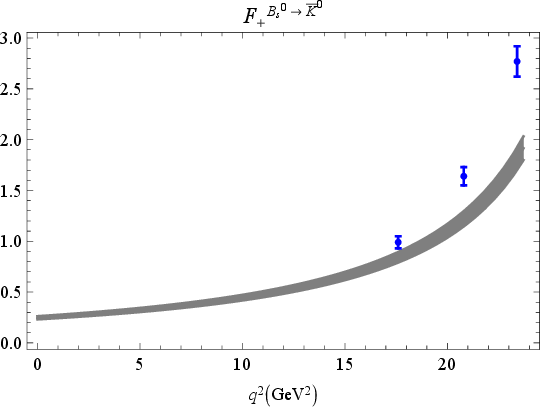}
\hfill\includegraphics[width=0.45\textwidth]{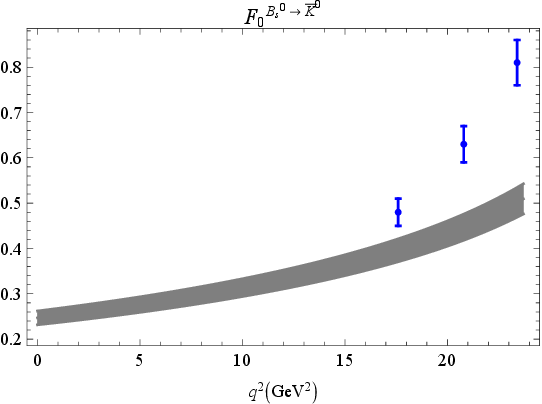}\\
\caption{$B \to \pi$ and $B_s \to \bar{K}^0$ form factors obtained in CCQM (solid broad lines) and in LQCD computations (dots with uncertainties) by RBC and UKQCD collaboration \cite{Flynn:2015mha}.}
\label{fig:lqcd_comparison}
\end{figure*}

\begin{table}
\caption{Partial branching fractions for $B^+ \to \pi^+ \mu^+ \mu^-$ in Unit of $10^{-9}$}
\begin{tabular*}{\textwidth}{@{\extracolsep{\fill}}lcccc@{}}
\hline\hline
$q^2$ bin & Nonresonant & Resonant & LQCD \cite{Bailey:2015nbd} & LCSR \cite{Hambrock:2015wka}\\
\hline
$[0.05,2.0]$	& $2.37 \pm 0.33$ & $3.62 \pm 0.43$ & -- 		& $2.49^{+0.30}_{-0.20}$\\
$[0.1,2.0]$		& $2.32 \pm 0.33$ & $3.58 \pm 0.42$ & 1.81 	& -- \\
$[2.0,4.0]$		& $2.50 \pm 0.35$ & $2.81 \pm 0.40$ & 1.92 	& $1.56^{+0.09}_{-0.08}$\\
$[4.0,6.0]$		& $2.54 \pm 0.35$ & $2.88 \pm 0.41$ & 1.91 	& $1.39^{+0.16}_{-0.11}$\\
$[6.0,8.0]$		& $2.65 \pm 0.37$ & $3.20 \pm 0.45$ & 1.89 	& $1.28^{+0.30}_{-0.23}$\\
$[15,17]$ 		& $1.48 \pm 0.21$ & $1.41 \pm 0.19$ & 1.69 	& --\\
$[17,19]$ 		& $1.20 \pm 0.17$ & $1.10 \pm 0.17$ & -- 		& --\\
$[19,22]$ 		& $1.23 \pm 0.17$ & $1.12 \pm 0.17$ & 1.84 	& --\\
$[22,25]$ 		& $0.52 \pm 0.07$ & $0.48 \pm 0.07$ & 1.07 	& --\\
$[1.0,6.0]$		& $6.28 \pm 0.89$ & $7.66 \pm 1.06$ & 4.78 	& $1.68^{+0.16}_{-0.12}$\\
$[15,22]$ 		& $3.90 \pm 0.54$ & $3.62 \pm 0.50$ & 5.05 	& --\\
$[4 m_\ell^2,q^2_{max}]$ & $21.73 \pm 3.04$ & -- 	& 20.4 	& --\\
\hline\hline
\end{tabular*}
\label{tab:br_Bpi_bins}
\end{table}
\begin{table*}
\caption{Branching fractions of $B^+$ and $B^0$ rare decays}
\begin{tabular*}{\textwidth}{@{\extracolsep{\fill}}l|lcccc@{}}
\hline\hline
Channel & Nonresonant & Resonant & LCSR \cite{Wu:2006rd} & RQM \cite{Faustov:2013pca,Faustov:2014zva} & Exp. \cite{Tanabashi:2018oca}\\
\hline
$10^8 \mathcal{B}(B^+ \to \pi^+ e^+ e^-)$			& $2.18 \pm 0.30$	& $1.82 \pm 0.18$ &  $1.89^{+0.23}_{-0.22}$ 	& -- & $< 8.0$\\
$10^8 \mathcal{B}(B^+ \to \pi^+ \mu^+\mu^-)$		& $2.17 \pm 0.30$	& $1.81 \pm 0.18$	& $1.88^{+0.24}_{-0.21}$	& $2.0 \pm 0.2$		 & $1.83 \pm 0.24 \pm 0.05$ \cite{Aaij:2015nea}\\
$10^8 \mathcal{B}(B^+ \to \pi^+ \tau^+\tau^-)$		& $0.53 \pm 0.15$	& $0.41 \pm 0.12$	& $0.90^{+0.13}_{-0.12}$	& $0.70 \pm 0.07$	 & --\\
$10^7 \mathcal{B}(B^+ \to \pi^+ \nu \bar{\nu})$	& $0.74 \pm 0.10$	& --	& -- & $1.2 \pm 0.1$		 	& $< 140$\\
\hline
$10^8 \mathcal{B}(B^0 \to \pi^0 e^+ e^-)$ 			& $1.01 \pm 0.14$	& $0.84 \pm 0.08$ & $0.87^{+0.11}_{-0.10}$ 	& -- & $< 8.4$\\
$10^8 \mathcal{B}(B^0 \to \pi^0 \mu^+ \mu^-)$ 	& $1.01 \pm 0.14$	& $0.84 \pm 0.08$ &$0.87^{+0.11}_{-0.10}$ 	& -- & $< 6.9$\\
$10^8 \mathcal{B}(B^0 \to \pi^0 \tau^+ \tau^-)$ 	& $0.24 \pm 0.07$	& $0.19 \pm 0.06$ & $0.41 \pm 0.06$ 			& -- & --\\
$10^7 \mathcal{B}(B^0 \to \pi^0 \nu \bar{\nu})$ 	& $0.34 \pm 0.05$	& -- & -- 									& --  & $< 90$\\
\hline
$10^8 \mathcal{B}(B^+\to \rho^+ e^+ e^-) $ 		& $4.82 \pm 2.39$	& $3.70 \pm 1.34$	& $4.0 \pm 0.4$ 		& -- & -- \\
$10^8 \mathcal{B}(B^+\to \rho^+ \mu^+\mu^-) $ 	& $4.05 \pm 1.45$	& $2.94 \pm 0.94$	& $3.9 \pm 0.4$			& $4.4 \pm 0.5$		 & -- \\
$10^8 \mathcal{B}(B^+\to \rho^+ \tau^+\tau^-) $ 	& $0.63\pm 0.14$	& $0.43 \pm 0.09$	& $0.40 \pm 0.04$		& $0.75 \pm 0.08$	 & --\\
$10^7 \mathcal{B}(B^+\to \rho^+ \nu\bar{\nu}) $ 	& $1.45 \pm 0.38$	& --	& -- &						$2.9\pm 0.3$		& $ <  300$\\
$10^7 \mathcal{B}(B^+ \to \rho^+ \gamma) $ 		& $8.55 \pm 1.38$	& --	& $13.8^{+1.6}_{-1.5}$& --			 & $9.8 \pm 2.5$\\
\hline
$10^8 \mathcal{B}(B^0 \to \rho^0 e^+ e^-)$ 		& $2.21 \pm 1.09$	& $1.70 \pm 0.61$ & $1.9 \pm 0.2$ & -- & --\\
$10^8 \mathcal{B}(B^0 \to \rho^0 \mu^+ \mu^-)$ 	& $1.86 \pm 0.66$	& $1.35 \pm 0.43$ & $1.8 \pm 0.2$ & -- & --\\
$10^8 \mathcal{B}(B^0 \to \rho^0 \tau^+ \tau^-)$ 	& $0.29 \pm 0.06$	& $0.20 \pm 0.04$ & $0.2 \pm 0.02$ & -- & --\\
$10^7 \mathcal{B}(B^0 \to \rho^0 \nu \bar{\nu})$ 	& $0.67 \pm 0.18$	& -- & -- & -- & $< 400$\\
$10^7 \mathcal{B}(B^0 \to \rho^0 \gamma) $ 		& $3.96 \pm 0.64$	& -- & $6.4 \pm 0.7$ & -- & --\\
\hline
$10^8 \mathcal{B}(B^0 \to \omega e^+ e^-)$ 		& $1.85 \pm 0.89$	& $1.41 \pm 0.59$ & $1.3 \pm 0.1$		& -- & --\\
$10^8 \mathcal{B}(B^0 \to \omega \mu^+ \mu^-)$& $1.57 \pm 0.55$	& $1.14 \pm 0.44$ & $1.2 \pm 0.1$	& -- & --\\
$10^8 \mathcal{B}(B^0 \to \omega \tau^+ \tau^-)$& $0.25 \pm 0.05$	& $0.18 \pm 0.03$ & $0.13 \pm 0.01$	& -- & --\\
$10^7 \mathcal{B}(B^0 \to \omega \nu \bar{\nu})$& $0.56 \pm 0.15$	& -- & -- & -- &$< 400$\\
$10^7 \mathcal{B}(B^0 \to \omega \gamma) $ 	& $3.11 \pm 0.50$	& -- & $5.8^{+0.6}_{-0.7}$ & -- & --\\
\hline\hline
\end{tabular*}
\label{tab:branching_B}
\end{table*}
\begin{table*}
\caption{Branching fractions of $B_s^0$ rare decays}
\begin{tabular*}{\textwidth}{@{\extracolsep{\fill}}l|lcccc@{}}
\hline\hline
Channel & Nonresonant & Resonant & LCSR \cite{Wu:2006rd}& RQM \cite{Faustov:2013pca,Faustov:2014zva}  & Exp.\\
\hline
$ 10^8 \mathcal{B}(B_s^0\to \bar{K}^*(892)^0 e^+ e^-)$			& $3.91 \pm 1.47$	& $2.95 \pm 0.80$	& $4.0 \pm 0.4$	& --		 & -- \\
$ 10^8 \mathcal{B}(B_s^0\to \bar{K}^*(892)^0\mu^+ \mu^-)$	& $3.32 \pm 0.94$	& $2.36 \pm 0.57$	& $3.8 \pm 0.3$	& $4.2 \pm 0.4$  & $2.9 \pm 1.0 \pm 0.2 \pm 0.3$ \cite{Aaij:2018jhg}\\
$ 10^8 \mathcal{B}(B_s^0\to \bar{K}^*(892)^0 \tau^+ \tau^-)$	& $0.53 \pm 0.10$	& $0.37 \pm 0.06$ & $0.50 \pm 0.04$& $0.75 \pm 0.08$	 & --\\
$ 10^7 \mathcal{B}(B_s^0\to \bar{K}^*(892)^0 \nu \bar{\nu})$	& $1.19 \pm 0.25$	& --	& --						& -- & -- \\
$10^7 \mathcal{B}(B_s^0\to \bar{K}^*(892)^0 \gamma)$ 			& $6.66 \pm 0.96$	& --	& $12.0^{+1.1}_{-1.2}$ & $3.0 \pm 0.3$		 & --\\
\hline
$10^8 \mathcal{B}(B_s^0 \to \bar{K}^0 e^+ e^-)$ 			& $1.65 \pm 0.21$	& $1.35 \pm 0.12$ & $1.99^{+0.21}_{-0.20}$	& --		 & --\\
$10^8 \mathcal{B}(B_s^0 \to \bar{K}^0 \mu^+ \mu^-)$ 	& $1.64 \pm 0.21$	& $1.35 \pm 0.12$	& $1.99^{+0.21}_{-0.20}$	& $2.2 \pm 0.2$		 & --\\
$10^8 \mathcal{B}(B_s^0 \to \bar{K}^0 \tau^+ \tau^-)$ 	& $0.39 \pm 0.09$	& $0.30 \pm 0.08$	& $0.74 \pm 0.07$ 			& $0.55 \pm 0.06$	& --\\
$10^7 \mathcal{B}(B_s^0 \to \bar{K}^0 \nu \bar{\nu})$ 	& $0.55 \pm 0.07$	& --	& -- & $1.41 \pm 0.14$	 	& --\\
\hline\hline
\end{tabular*}
\label{tab:branching_Bs}
\end{table*}
\begin{table}
\caption{Ratios of the Branching fractions}
\begin{tabular*}{\textwidth}{@{\extracolsep{\fill}}lccc@{}}
\hline\hline
Ratio & Unit & Present & Data \\
\hline
$\frac{\mathcal{B} (B^+ \to \pi^+ \mu^+ \mu^-)}{\mathcal{B} (B^+ \to K^+ \mu^+ \mu^-)}$ 	& $10^{-2}$ & $3.02 \pm 0.42$  & $5.3 \pm 1.4 \pm 0.1$  \cite{LHCb:2012de}\\
$\frac{\mathcal{B} (B^+ \to \rho^+ \mu^+ \mu^-)}{\mathcal{B} (B^+ \to K^{*}(892)^+ \mu^+ \mu^-)}$ & $10^{-2}$	& $3.19 \pm 1.14$  & --  \\
$\frac{\mathcal{B}(B_s^0\to \bar{K}^*(892)^0\mu^+ \mu^-)}{\mathcal{B}(\bar{B}^0\to \bar{K}^*(892)^0\mu^+ \mu^-)}$ & $10^{-2}$ & $2.61 \pm 0.74$ & $3.3 \pm 1.1 \pm 0.3 \pm 0.2$ \cite{Aaij:2018jhg}\\
\hline\hline
\end{tabular*}
\label{tab:branching_ratio}
\end{table}

\begin{figure*}[htbp]
\includegraphics[width=0.45\textwidth]{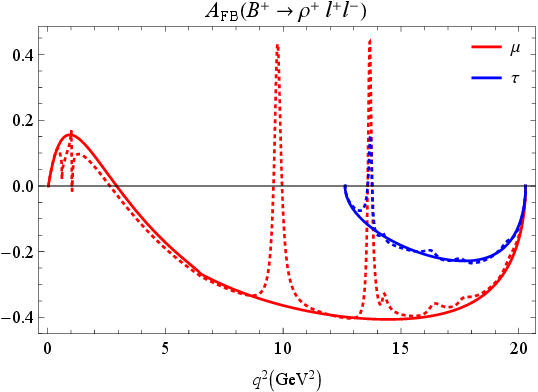}
\hfill\includegraphics[width=0.45\textwidth]{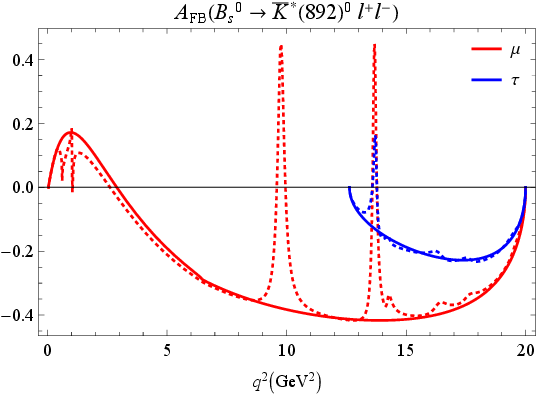}\\
\vspace{0.25cm}
\includegraphics[width=0.45\textwidth]{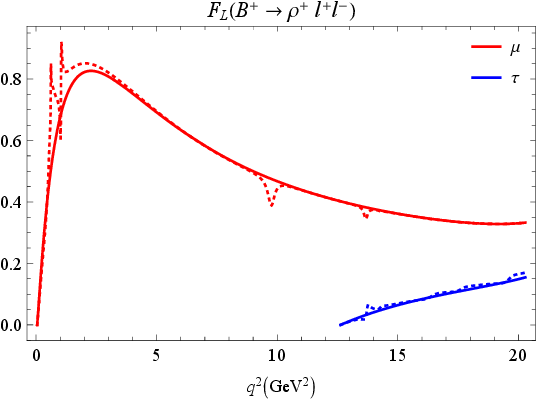}
\hfill\includegraphics[width=0.45\textwidth]{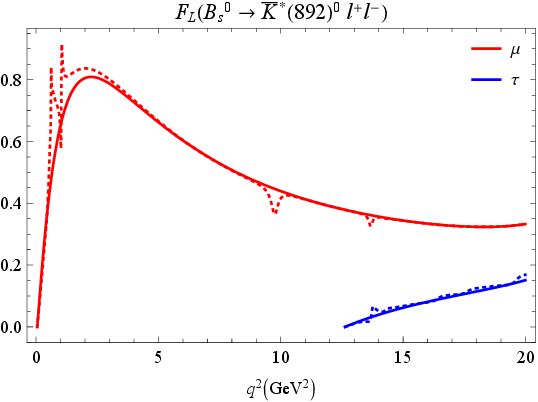}\\
\vspace{0.25cm}
\includegraphics[width=0.45\textwidth]{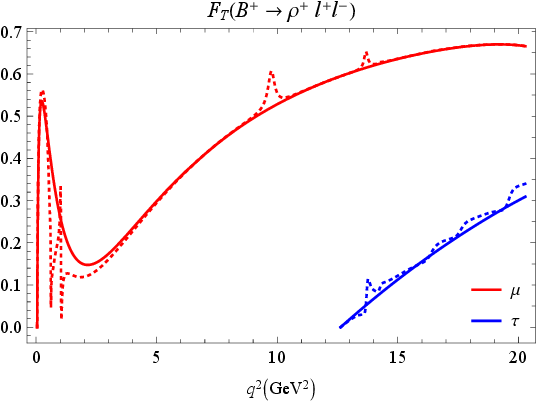}
\hfill\includegraphics[width=0.45\textwidth]{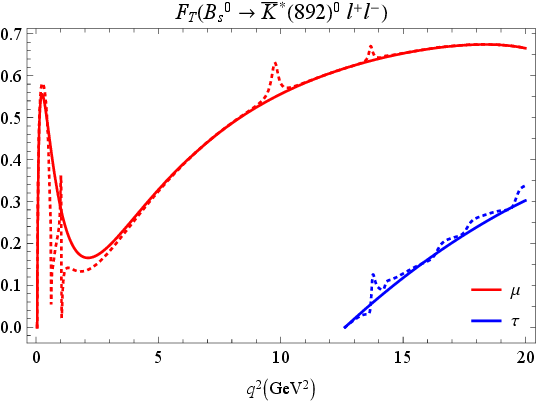}\\
\caption{Forward backward asymmetry, longitudinal and transverse polarizations (solid lines - excluding resonances, dashed lines - including vector resonances). }
\label{fig:asymmetry}
\end{figure*}

\begin{figure*}[htbp]
\includegraphics[width=0.45\textwidth]{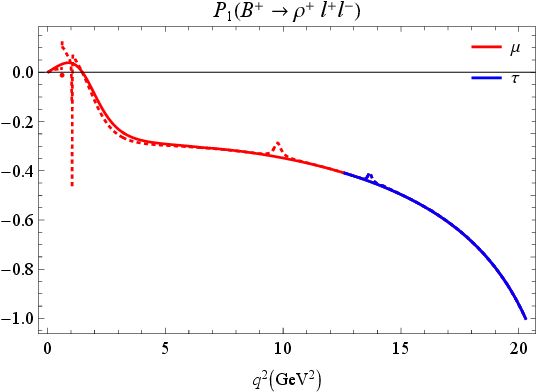}
\hfill\includegraphics[width=0.45\textwidth]{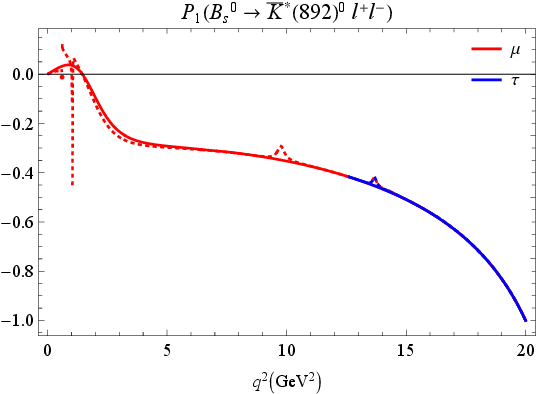}\\
\vspace{0.25cm}
\includegraphics[width=0.45\textwidth]{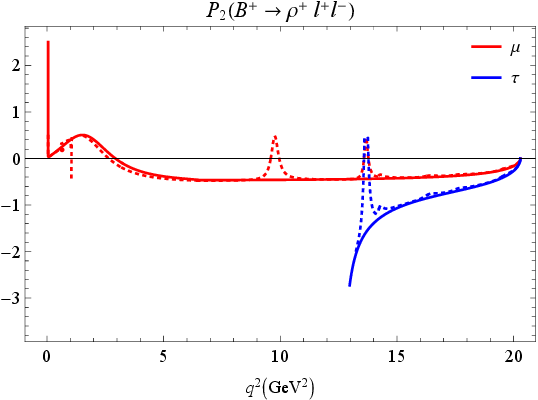}
\hfill\includegraphics[width=0.45\textwidth]{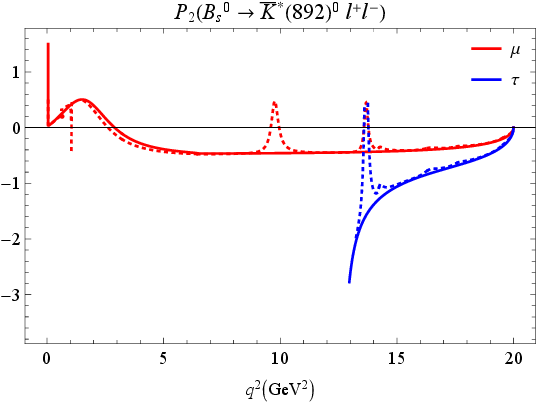}\\
\vspace{0.25cm}
\includegraphics[width=0.45\textwidth]{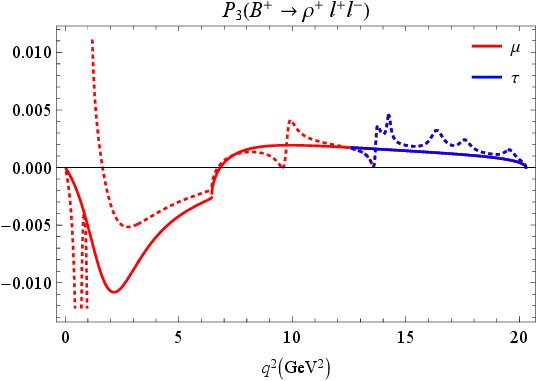}
\hfill\includegraphics[width=0.45\textwidth]{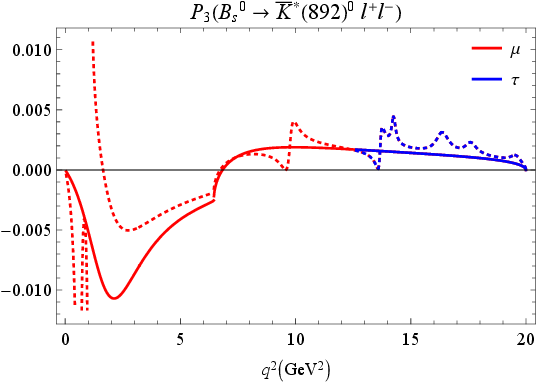}\\
\caption{Clean observables $P_{1,2,3}$ in whole $q^2$ range (solid lines - excluding resonances, dashed lines - including vector resonances).}
\label{fig:p123}
\end{figure*}

\begin{figure*}[htbp]
\includegraphics[width=0.45\textwidth]{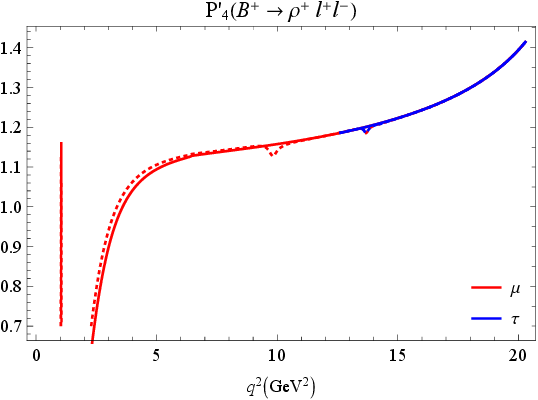}
\hfill\includegraphics[width=0.45\textwidth]{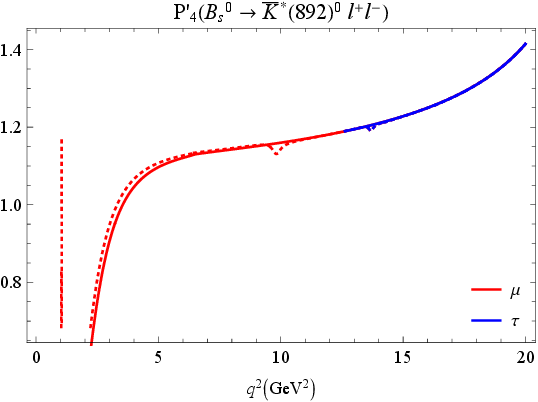}\\
\vspace{0.25cm}
\includegraphics[width=0.45\textwidth]{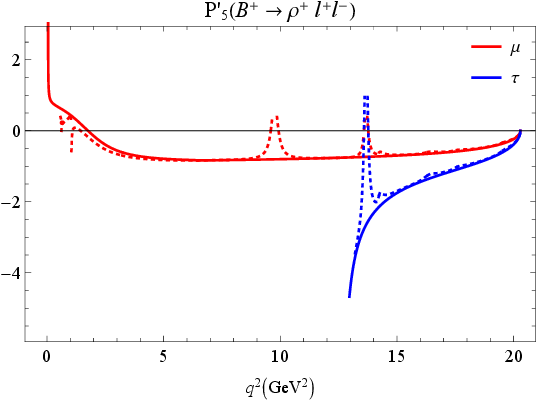}
\hfill\includegraphics[width=0.45\textwidth]{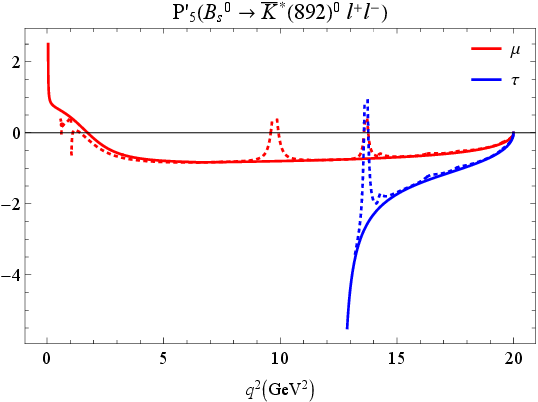}\\
\vspace{0.25cm}
\includegraphics[width=0.45\textwidth]{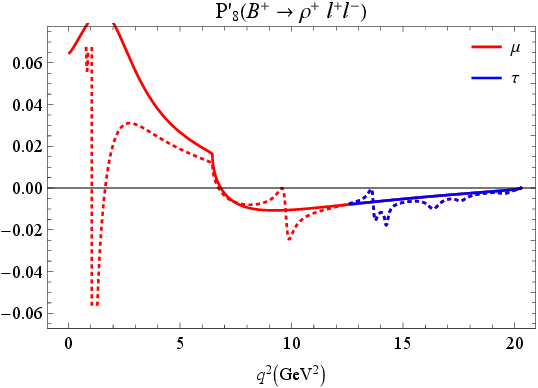}
\hfill\includegraphics[width=0.45\textwidth]{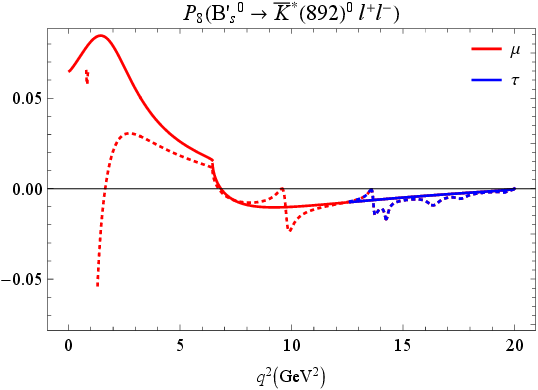}\\
\caption{Clean observables $P^\prime_{4,5,8}$ in whole $q^2$ range (solid lines - excluding resonances, dashed lines - including vector resonances).}
\label{fig:p458}
\end{figure*}

\begin{table*}
\caption{$q^2$- averages of polarization observables over the whole allowed kinematic region for $B^+ \to \rho^+ \ell^+ \ell^-$, $B^0 \to (\rho^0, \omega) \ell^+ \ell^-$ and $B_s^0 \to \bar{K}^*(892)^0 \ell^+ \ell^-$.}
\begin{tabular*}{\textwidth}{@{\extracolsep{\fill}}r|ccc|ccc|ccc|ccc@{}}
\hline\hline
Obs. & \multicolumn{3}{c|}{$B^+ \to \rho^+ \ell^+ \ell^-$} & \multicolumn{3}{c|}{$B^0 \to \rho^0 \ell^+ \ell^-$} & \multicolumn{3}{c|}{$B^0 \to \omega \ell^+ \ell^-$} & \multicolumn{3}{c}{$B_s^0\to \bar{K}^*(892)\ell^+ \ell^-$} \\
& $e^+ e^-$ & $\mu^+ \mu^-$ & $\tau^+ \tau^-$ & $e^+ e^-$ & $\mu^+ \mu^-$ & $\tau^+ \tau^-$ & $e^+ e^-$ & $\mu^+ \mu^-$ & $\tau^+ \tau^-$ & $e^+ e^-$ & $\mu^+ \mu^-$ & $\tau^+ \tau^-$\\
\hline
$-\langle A_{FB} \rangle$		
& 0.178	& 0.226 & 0.207 	
& 0.179 & 0.227 & 0.209
& 0.184 & 0.229 & 0.203
& 0.194 & 0.244 & 0.207\\
$\langle F_L\rangle$				
& 0.417 & 0.510 & 0.110 	
& 0.414 & 0.506 & 0.110
& 0.415 & 0.501 & 0.111
& 0.394 & 0.479 & 0.107\\
$\langle F_T\rangle$				
& 0.563 & 0.455 & 0.217 	
& 0.567 & 0.458 & 0.218
& 0.566 & 0.465 & 0.218
& 0.588 & 0.488 & 0.218\\
$-\langle P_1 \rangle$			
& 0.297	& 0.460 & 0.688		
& 0.294 & 0.457 & 0.686
& 0.322 & 0.486 & 0.706
& 0.317 & 0.475 & 0.701\\
$-\langle P_2 \rangle$			
& 0.211 & 0.331 & 0.636	
& 0.211 & 0.331 & 0.637
& 0.216 & 0.328 & 0.620
& 0.220 & 0.333 & 0.631\\
$10^{4} \times \langle P_3 \rangle$	
& 0.326 & 2.198 & 17.694	
& $-$1.422 & $-$0.518 & 13.063
& 0.526 & 2.498 & 17.584
& 0.930 & 2.934 & 17.175\\
$\langle P_4^\prime \rangle$	
& 0.735 & 0.993 & 1.297	
& 0.738 & 0.994 & 1.296
& 0.768 & 1.017 & 1.304
& 0.776 & 1.019 & 1.303\\
$-\langle P_5^\prime \rangle$	
& 0.385 & 0.543 & 1.000 	
& 0.384 & 0.543 & 1.002
& 0.388 & 0.537 & 0.970
& 0.401 & 0.549 & 0.985\\
$10^{2} \times \langle P_8^\prime \rangle$	
& 3.058 & 2.741 & $-$0.468
& 3.112 & 2.802 & $-$0.372
& 2.905 & 2.559 & $-$0.447
& 2.826 & 2.475 & $-$0.440\\
$-\langle S_3 \rangle$			
& 0.083	& 0.105 & 0.074	
& 0.083 & 0.105	 & 0.075
& 0.091 & 0.113 & 0.077
& 0.093 & 0.116 & 0.077\\
$\langle S_4 \rangle$				
& 0.178	& 0.239 & 0.100 	
& 0.179 & 0.239 & 0.101
& 0.186 & 0.246 & 0.102
& 0.186 & 0.246 & 0.100\\
\hline\hline
\end{tabular*}
\label{tab:obs_BBs}
\end{table*}

\begin{table*}
\caption{Binned value of observables for $B \to \rho \ell^+ \ell^-$ for $\ell = e$ and $\mu$}
\begin{tabular*}{\textwidth}{@{\extracolsep{\fill}}cc|cc|cc|c@{}}
\hline\hline
Obs.& $q^2$& \multicolumn{2}{c|}{$B \to \rho e^+e^-$} & \multicolumn{3}{c}{$B \to \rho \mu^+\mu^-$}\\ \cline{3-7}
&& $B^+ \to \rho^+$ &  $B^0 \to \rho^0$ & $B^+ \to \rho^+$ &  $B^0 \to \rho^0$ & LCSR \cite{Kindra:2018ayz} \\
\hline
$10^9 \times \mathcal{B}$ &$[0.1, 0.98]$ & $2.58 \pm 2.18$ & $1.19 \pm 0.98$ &  $2.54 \pm 2.15$ & $1.17 \pm 0.97$ & $2.165 \pm 0.302$\\
&$[1.1, 6]$ & $7.51 \pm 4.52$ & $3.42 \pm 2.07$ & $7.49 \pm 4.52$ & $3.41 \pm 2.07$ & $4.064 \pm 0.778$\\
\hline
$\langle A_{FB} \rangle$ &$[0.1, 0.98]$ & 0.089 & 0.090 & 0.079 & 0.080 & $-0.046 \pm 0.005$\\
&$[1.1, 6]$ & $-0.085$ & $-0.085$ & $-0.085$ & $-0.085$ & $-0.024 \pm 0.018$\\
\hline
$\langle R_{\rho} \rangle$ &$[0.1, 0.98]$ & 0.984  & 0.983 & 0.984 & 0.983 & $ 0.955 \pm 0.194$\\
&$[1.1, 6]$ & 0.997 & 0.997 & 0.997 & 0.997 & $1.036 \pm 0.289$\\
\hline
$\langle F_L^{\rho} \rangle$ &$[0.1, 0.98]$ & 0.511 & 0.506 & 0.447 & 0.443 & $0.409 \pm 0.067$\\
&$[1.1, 6]$ & 0.779 & 0.776 & 0.763 & 0.761 & $0.822 \pm 0.039$\\
\hline
$\langle P_1 \rangle$ &$[0.1, 0.98]$ & 0.017 & 0.016 & 0.019 & 0.018 & $0.050 \pm 0.181$\\
&$[1.1, 6]$ & $-0.254$ & $-0.252$ & $-0.255$ & $-0.253$ & $-0.044 \pm 0.110$\\
\hline
$\langle P_2 \rangle$ &$[0.1, 0.98]$ & 0.122 & 0.121 & 0.146 & 0.146 & $0.083 \pm 0.010$\\
&$[1.1, 6]$ & $-0.256$ & $-0.254$ & $-0.259$ & $-0.257$ & $0.074 \pm 0.053$\\
\hline
$\langle P_3 \rangle$ &$[0.1, 0.98]$ & $-$0.008 & $-$0.008 & $-$0.009 & $-$0.009 & $-0.228 \pm 0.044$\\
&$[1.1, 6]$ & $-$0.003 & $-$0.003& $-$0.003  & $-$0.003 & $-0.229 \pm 0.028$\\
\hline
$\langle P_4^\prime \rangle$ &$[0.1, 0.98]$ & $-0.291$ & $-0.282$ & $-0.288$ & $-0.279$ & $-0.591 \pm 0.077$\\
&$[1.1, 6]$ & 0.878 & 0.879 & 0.882 & 0.883 & $0.470 \pm 0.161$\\
\hline
$\langle P_5^\prime \rangle$ &$[0.1, 0.98]$ & 0.447 & 0.453 & 0.468 & 0.474 & $0.368 \pm 0.043$\\
&$[1.1, 6]$ & $-0.625$ & $-0.622$ & $-0.632$ & $-0.628$ & $-0.178 \pm 0.084$\\
\hline
$\langle P_8^\prime \rangle$ &$[0.1, 0.98]$ & 0.312 & 0.313 & 0.321 & 0.322 & $-0.133 \pm 0.021$\\
&$[1.1, 6]$ & 0.014 & 0.014  & 0.014 & 0.014 & $0.113 \pm 0.013$\\
\hline\hline
\end{tabular*}
\label{tab:obs_B_bin}
\end{table*}

\begin{table*}
\caption{Binned value of observables for $B^0 \to \omega \ell^+ \ell^-$ for $\ell = e$ and $\mu$}
\begin{tabular*}{\textwidth}{@{\extracolsep{\fill}}cc|cc||cc|cc@{}}
\hline\hline
Obs. & $q^2$ & $B^0 \to \omega e^+ e^-$ & $B^0 \to \omega \mu^+ \mu^-$ & Obs. & $q^2$ & $B^0 \to \omega e^+ e^-$ & $B^0 \to \omega \mu^+ \mu^-$\\
\hline
$10^9 \times \mathcal{B}$ &$[0.1, 0.98]$ & $0.94 \pm 0.80$ & $0.92 \pm 0.78$
& $\langle P_2 \rangle$ & $[0.1, 0.98]$ & 0.123 & 0.147\\
&$[1.1, 6]$ & $2.78 \pm 1.67$ & $2.77 \pm 1.66$
& & $[1.1, 6]$ & $-0.259$ & $-0.263$\\
\hline
$\langle A_{FB} \rangle$ &$[0.1, 0.98]$ & 0.090 & 0.080
& $\langle P_3 \rangle$ & $[0.1, 0.98]$ & $-$0.009 & $-$0.010\\
&$[1.1, 6]$ & $-$0.088 & $-$0.087
& & $[1.1, 6]$ & $-$0.003 & $-$0.003\\
\hline
$\langle R_{\omega} \rangle$ & $[0.1, 0.98]$ & 0.979 & 0.979
& $\langle P_4^\prime \rangle$ &$[0.1, 0.98]$ & $-0.286$ & $-0.282$\\
&$[1.1, 6]$ & 0.996 & 0.996
& & $[1.1, 6]$ & 0.894 & 0.898\\
\hline
$\langle F_L^{\omega} \rangle$ & $[0.1, 0.98]$ & 0.511 & 0.447
& $\langle P_5^\prime \rangle$ &$[0.1, 0.98]$ & 0.447 & 0.469\\
&$[1.1, 6]$ & 0.774 & 0.759
& & $[1.1, 6]$ & $-0.618$ & $-0.625$\\
\hline
$\langle P_1 \rangle$ & $[0.1, 0.98]$ & 0.018 & 0.021
& $\langle P_8^\prime \rangle$ &$[0.1, 0.98]$ & 0.312 & 0.321\\
&$[1.1, 6]$ & $-0.273$ & $-0.274$
& & $[1.1, 6]$ & 0.014 & 0.014\\
\hline\hline
\end{tabular*}
\label{tab:obs_B0_bin}
\end{table*}

\begin{table*}
\caption{Binned value of observables for $B_s^0 \to \bar{K}^*(892)^0 \ell^+ \ell^-$ for $\ell = e$ and $\mu$}
\begin{tabular*}{\textwidth}{@{\extracolsep{\fill}}cc|cc|ccc@{}}
\hline\hline
Obs. & $q^2$& \multicolumn{2}{c|}{$B_s^0\to \bar{K}^*(892)^0e^+e^-$} & \multicolumn{3}{c}{$B_s^0\to \bar{K}^*(892)^0\mu^+\mu^-$}\\
&& Present & pQCD \cite{Jin:2020jtu} &  Present & pQCD \cite{Jin:2020jtu} & LCSR \cite{Kindra:2018ayz}\\
\hline
$10^9 \times \mathcal{B}$ &$[0.1, 0.98]$ & $1.89 \pm 1.25$ & $1.63^{+0.65}_{-0.43}$ & $1.86 \pm 1.23$ & $1.60^{+0.64}_{-0.42}$ & $3.812 \pm 0.450 \pm 0.086$\\
&$[1.1, 6]$ & $5.54 \pm 2.73$ & $5.22^{+2.13}_{-1.59}$ & $5.51 \pm 2.72$ & $5.21^{+2.12}_{-1.58}$ & $7.803 \pm 1.758 \pm 0.357$\\
\hline
$\langle A_{FB} \rangle$ &$[0.1, 0.98]$ & 0.096 & 0.096 (2) & 0.085 & 0.085 (2) & $-0.060 \pm 0.008 \pm 0.001$\\
&$[1.1, 6]$ & $-0.099$ & $-0.064 (3)$ & $-0.098$ & $-0.064 (3)$ & $-0.029 \pm 0.020 \pm 0.009$\\
\hline
$\langle R_{K^*} \rangle$ &$[0.1, 0.98]$ & 0.984 & 0.984 (1) & 0.984 & 0.984 (1) & $0.940 \pm 0.009 \pm$ 0.001\\
&$[1.1, 6]$ & 0.995 & 0.997 (1) & 0.995 & 0.997 (1) & $0.998 \pm 0.004 \pm 0.0$ \\
\hline
$\langle F_L^{K^*} \rangle$ &$[0.1, 0.98]$ & 0.479 & 0.442 (8) & 0.420 & 0.446 (8) & $0.453 \pm 0.067 \pm$ 0.014\\
&$[1.1, 6]$ & 0.752 & 0.780 (10) & 0.738 & 0.783 (10) & $0.853 \pm 0.038 \pm 0.007$ \\
\hline
$\langle P_1 \rangle$ &$[0.1, 0.98]$ & 0.016 & 0.110 (2) & 0.019 & 0.012 (2) & $0.012 \pm 0.129 \pm 0.001$\\
&$[1.1, 6]$ & $-0.257$ & $-0.306 (30)$ & $-0.258$ & $-0.307 (30)$ & $-0.081 \pm 0.111 \pm 0.005$\\
\hline
$\langle P_2 \rangle$ &$[0.1, 0.98]$ & 0.123 & 0.115 (1) & 0.147 & 0.129 (1) & $0.118 \pm 0.013 \pm 0.001$\\
&$[1.1, 6]$ & $-0.266$ & $-0.193 (10)$ & $0.270$ & $-0.196 (10)$ & $0.112 \pm 0.072 \pm 0.036$\\
\hline
$\langle P_3 \rangle$ &$[0.1, 0.98]$ & $-$0.008 & 0.005 (1) & $-$0.009 & 0.005 (1) & $0.001 \pm 0.002 \pm 0.0$\\
&$[1.1, 6]$ & $-$0.003 & 0.002 (1) & $-$0.003 & 0.002 (1) & $0.004 \pm 0.010 \pm 0.002$\\
\hline
$\langle P_4^\prime \rangle$ &$[0.1, 0.98]$ & $-0.284$ & $-0.369 (3)$ & $-0.280$ & $-0.365 (3)$ & $-0.593 \pm 0.057 \pm 0.009$\\
&$[1.1, 6]$ & 0.890 & 0.895 (10) & 0.894 & 0.899 (10) & $0.464 \pm 0.164 \pm 0.014$\\
\hline
$\langle P_5^\prime \rangle$ &$[0.1, 0.98]$ & 0.445 & 0.541 (1) & 0.466 & 0.512 (1) & $0.547 \pm 0.051 \pm 0.016$\\
&$[1.1, 6]$ & $-0.634$ & $-0.575 (40)$ & $-0.641$ & $-0.578 (40)$ & $-0.286 \pm 0.125 \pm 0.046$\\
\hline
$\langle P_8^\prime \rangle$ &$[0.1, 0.98]$ & 0.313 & 0.396 (1) & 0.322 & 0.399 (1) & $0.015 \pm 0.003 \pm 0.016$\\
&$[1.1, 6]$ & 0.014 & 0.006 (1) & 0.014 & 0.006 (1) & $0.040 \pm 0.004 \pm 0.017$\\
\hline\hline
\end{tabular*}
\label{tab:obs_Bs_bin}
\end{table*}

In Tab. \ref{tab:ff_comparison_BP} - \ref{tab:ff_comparison_BsKv}, we provide the brief comparison of the form factors at the maximum recoil $q^2 = 0$ with the light cone sum rules (LCSR), perturbative QCD (pQCD), supersymmetry (SUSY), relativistic quark model (RQM), soft collinear effective field theory (SCET), constituent quark models (CQM) and light front quark model (LFQM).
For $B \to \pi$ channel, our form factors are in very good agreement with LCSR \cite{Lu:2018cfc,Wu:2006rd}, perturbative QCD \cite{Wang:2012ab} and CQM \cite{Melikhov:2000yu} where as for $B_s^0 \to \bar{K}^0$ channel, our form factors underestimate the LCSR and CQM results but are closer with the pQCD prediction \cite{Jin:2020jtu}. Note that in pQCD Ref. \cite{Jin:2020jtu}, the form factors are computed in the framework of perturbative QCD using the inputs from LQCD.
Similarly, in Tab. \ref{tab:ff_comparison_Brho}, \ref{tab:ff_comparison_Bomega} and \ref{tab:ff_comparison_BsKv} we present $B \to \rho$,$B^0 \to \omega$ and $B_s^0 \to \bar{K}^*(982)^0$ transition form factors and our results are matching well with the LCSR results. For $B_s^0 \to \bar{K}^*(982)^0$ channel, our form factors are also in good aggreement with the pQCD approach \cite{Jin:2020jtu}.

Further, we also compare the form factors in the entire dynamical range of momentum transfer $0 \leq q^2 \leq q^2_{\mathrm{max}} = (m_{B_{(s)}} - m_{P/V})^2$. In Fig. \ref{fig:B_P} - \ref{fig:B_V_Ts}, we provide the $q^2$ dependency of form factors with comparison to different approaches.
In Fig. \ref{fig:B_P}, we provide the form factor comparison of $B \to \pi$ and $B_s^0 \to \bar{K}^0$ channels along with different theoretical approaches. Note that the form factor $F_0(q^2)$ is related with $F_+(q^2)$ and $F_-(q^2)$ of Eq. (\ref{eq:ff_PP}) via relation
\begin{eqnarray}
F_0(q^2) & = & F_+(q^2) + \frac{q^2}{m_1^2 - m_2^2} F_-(q^2)
\label{eq:ff_f0}
\end{eqnarray}
It is worth mentioning that our results are matching fairly well with other approaches for $q^2 \leq 15$ GeV$^2$ for both the channels. It is also observed that our form factors $F_{+,0} (q^2)$ are in close resemblance with those obtained using LFQM \cite{Verma:2011yw, Chang:2019mmh} and our form factors $F_{0,T} (q^2)$ are also in close resembles with CQM \cite{Melikhov:2000yu}.

In Figs. \ref{fig:B_V} $-$ \ref{fig:B_V_Ts}, we provide the form factor comparison for the channels $B \to \rho$, $B^0 \to \omega$ and $B_s^0 \to \bar{K}^*(892)^0$ and it is observed that for the form factors $V(q^2)$ and $A_{0,1,2} (q^2)$ show good agreement with other theoretical predictions for the whole $q^2$ range. Our results are also matching well for the form factors $T_{1,2,3} (q^2)$ with other approaches.

In Tab. \ref{tab:PP_FF_LQCD}, we compare our results of form factors with LQCD from RBC and UKQCD collaborations \cite{Flynn:2015mha} at higher $q^2$ values for the channels $B \to \pi$ and $B_s^0 \to \bar{K}^{0}$. For intermediate $q^2$ range, our results are closer to the LQCD predictions but for $q^2 \to q^2_{\mathrm{max}}$, they are systematically lower.
In Ref. \cite{Flynn:2015mha}, the vector and scalar form factors for $B \to \pi \ell \nu_\ell$ and $B_s \to K \ell \nu_\ell$ are computed in LQCD at three $q^2$ ranges using domain-wall light quarks and relativistic b-quarks.
Similar trend is also observed for $B_s^0 \to \bar{K}^*(892)^0$ (Tab. \ref{tab:PV_FF_LQCD}). It is interesting to note that similar nature was also observed for $D \to (\pi, K)$ form factors in \cite{Soni:2017eug,Soni:2018adu,Ivanov:2019nqd} when compared with LQCD predictions from ETM collaboration. However, the tensor form factor shows very good agreement with the ETM collaboration.
In Fig. \ref{fig:lqcd_comparison}, we show the form factors for the channels $B \to \pi$ and $B_s^0 \to \bar{K}^0$ in comparison with LQCD predictions \cite{Flynn:2015mha}. We also present the spread of our form factors in the whole $q^2$ range because of the uncertainties in the fitting parameters. Similar spread can also be obtained for the vector meson form factors.

Utilizing the model dependent and independent parameters along with the form factors, we compute the branching fractions using Eq. \ref{eq:branching}.
We present our results with and without resonant counterparts from the charmed ($J/\psi$ and $\psi (2S)$) and charmless ($\rho, \omega$ and $\phi$) vector meson resonances.
For computation of resonant branching fractions, the $q^2$ range close to the $J/\psi$ and $\psi (2S)$ are avoided as experimental analysis also excludes these resonance regions.
The experimentally vetoed regions corresponding to $J/\psi$ and $\psi (2S)$ are $8.0 < q^2 < 11.0$ GeV$^2$ and $12.5 < q^2 < 15.0$ GeV$^2$, respectively.
It is important to note here that when we consider the resonances for branching fraction computations, the results are enhanced by an order or two. Similar observations have also been reported in a review \cite{Blake:2016olu}.
Further, in order to compare our results with experimental data, we also exclude the experimentally vetoed regions.
In the literature, there are different ways for the treatment of these resonance regions.
For example, in Ref. \cite{Ali:2013zfa}, the authors have smoothed-out the resonance regions by incorporating the next-to-leading order correction in $C_9^{\mathrm{eff}}$ using the auxiliary functions $F_{1,2}^{(7,9)} (q^2)$ from the ref. \cite{Seidel:2004jh}.
In Tab. \ref{tab:br_Bpi_bins}, we compare our results of branching fractions for $B^+ \to \pi^+ \mu^+\mu^-$ in narrow $q^2$ bins with LQCD and LCSR results and it is observed that our results are on higher side. We also note that the results from most of the theory attempts are systematically higher than LHCb data in the narrow $q^2$ bins \cite{Aaij:2015mea}.
In Tab. \ref{tab:branching_B}, we present our results of rare decays of $B$ and $B^0$ mesons in comparison with theoretical approaches viz. LCSR \cite{Wu:2006rd} and RQM \cite{Faustov:2013ima}.
Note that the results presented in Refs. \cite{Wu:2006rd,Faustov:2013ima} corresponds to the nonresonant contributions.
On the experimental front, LHCb collaboration has provided the branching fractions for $B^+ \to \pi^+ \mu^+ \mu^-$ and our resonant result is lower than LHCb data. For the other channels, only the upper bounds are provided in PDG.
Further, Belle \cite{Wei:2008nv} and \textit{BABAR} \cite{Aubert:2007mm,Lees:2013lvs} collaborations have also performed the search for $B^0 \to \pi^0 \ell^+ \ell^-$ with $\ell = e$ and $\mu$ and our results are well within their upper limit.
For $B^+ \to \rho^+ \ell^+ \ell^-$ and $B^0 \to (\rho^0, \omega) \ell^+ \ell^-$ channels, the experimental data is yet to be reported. For the $B^{+(0)} \to (\rho^{+(0)}, \omega) \nu \bar{\nu}$ channels, our results are well within the upper limit of PDG data.
Our results are within the range predicted in the LHCb data for the channel $B_s^0 \to \bar{K}^*(892)^0 \mu^+ \mu^-$ (Tab. \ref{tab:branching_Bs}) and for the other rare $B_s$ decays, we again do not have experimental results available.
The ratio for muon channel to electron channel for $B^+ \to \pi^+$ and $B^0 \to \pi^0$ tends to be $1$, whereas for the $B^{+(0)} \to (\rho^{+(0)}, \omega)$, the ratio in our study comes out to be 0.85. However, considering the uncertainties in form factors and transporting them to corresponding branching fractions, the ratio approaches unity.
It is also observed that the nonresonant branching fractions corresponding to $B^+ \to (\pi^+,\rho^+) \ell^+ \ell^-$ for $\ell = e,~\mu$ are in good agreement with the LCSR and $\mu$ channel of RQM results, whereas for $B_s^0$ channels, our results are systematically lower than LCSR and RQM results.
Our results are in good agreement within the uncertainties of LCSR for the $B^0 \to (\pi^0, \rho^0, \omega) \ell^+ \ell^-$ and $B^+ \to \rho^+ \tau^+ \tau^-$ channels.
We also compute the radiative decays and it is observed that our results on $\mathcal{B} (B^+ \to \rho^+ \gamma)$ are within the uncertainty range reported by LHCb. Our results on $\mathcal{B} (B_s^0 \to \bar{K}^*(892)^{0} \gamma)$ are within the uncertainty limits of LHCb data but disagree with the RQM and LCSR results.

CCQM was also employed for studying the rare decays corresponding the $b \to s$ transitions in Ref.\cite{Dubnicka:2015iwg,Dubnicka:2016nyy} by Dubni\v{c}ka \textit{et al}. In these articles they have computed the branching fractions for $B \to K^{(*)} \ell^+ \ell^-$ and $B_s \to \phi \ell^+ \ell^-$ for $\ell = e, \mu, \tau$.
Using the inputs from these papers, we compute the ratios of $b \to s$ to $b \to d$ rare decays and they are tabulated in Tab. \ref{tab:branching_ratio}.
It is worth mentioning that our ratios are well within the range predicted in the LHCb data except for the ratio $\mathcal{B} (B^+ \to \pi^+ \mu^+ \mu^-)/\mathcal{B} (B^+ \to K^+ \mu^+ \mu^-)$ where our result underestimate the LHCb data.

We also compute some more physical observables defined in terms of helicity structure functions and that can be measured experimentally. These observables are helpful in understanding the influence of different flavor of leptons in the final stage that is described in terms of angular distributions between the charged leptons pairs and momentum of daughter meson.
This distribution allows us to compute the different observables such as decay widths, forward backward asymmetry and polarization of daughter mesons. The detailed computation technique for computation of these observables are available for the rare decays of $B_{(c)}$ mesons in \cite{Faessler:2002ut,Dubnicka:2015iwg}. We employ the same formalism for present study. These observables are explicitly expressed in terms of helicity structure functions and form factors. The forward-backward asymmetry defined as
\begin{eqnarray}
A_{\mathrm{FB}}^\ell = \frac{1}{d\Gamma/dq^2} \left(\int_0^1 - \int_{-1}^0\right) d\cos \theta \frac{d^2\Gamma}{dq^2 d\cos \theta} = \frac34 \beta_\ell \frac{\mathcal{H}_P^{12}}{\mathcal{H}_{\mathrm{tot}}}.
\label{eq:asymmetry}
\end{eqnarray}
Similarly, longitudinal and transverse polarization is defined as
\begin{eqnarray}
F_L  =  \frac12 \beta_\ell^2 \frac{\mathcal{H}_L^{11} + \mathcal{H}_L^{22}}{\mathcal{H}_{\mathrm{tot}}}, \qquad F_T  =  \frac12 \beta_\ell^2 \frac{\mathcal{H}_U^{11} + \mathcal{H}_U^{22}}{\mathcal{H}_{\mathrm{tot}}}.
\end{eqnarray}
In Eq. (\ref{eq:asymmetry}), the angle $\theta$ is the polar angle between the momentum transfer ($\vec{q} = \vec{p}_1 -\vec{p}_2$) and momentum of parent meson ($\vec{k}_1$) in the $\ell^+\ell^-$ rest frame. The computation of $q^2$ averages value of these observables can be done by multiplying by phase factor $|{\bf p_2}| q^2 (1 - m_\ell^2/q^2)^2$ in the numerator and denominator explicitly.
In Fig. \ref{fig:asymmetry}, we plot forward-backward asymmetry, longitudinal polarization and transverse polarization in the whole $q^2$ range considering both resonant and nonresonant contributions. We also provide the averages of these resonant observables in Tab. \ref{tab:obs_BBs}. Note here that for computing the resonant contribution of these observables, we exclude the experimental vetoed regions corresponding to vector resonances.

The helicity amplitudes corresponding to form factors are also utilised for further computations of so-called clean observables \cite{Matias:2012xw}.
These observables have been reported by LHCb \cite{Aaij:2015oid,Aaij:2020nrf} and Belle \cite{Wehle:2016yoi} collaboration for $b \to s$ rare decay and similarly these observables are also expected for the $b \to d$ decays.
We compute these observables $P_i$ using the relations expressed in Refs. \cite{Dubnicka:2015iwg,DescotesGenon:2012zf}.
In Fig. \ref{fig:p123} and \ref{fig:p458}, we display the plots for the clean observables $P_{1,2,3}$ and $P^\prime_{4,5,8}$ for the channels $B^+ \to \rho^+$ and $B_s^0 \to \bar{K}^*(892)^{0}$. Similar plots can also obtained for the $B^0 \to (\rho^0, \omega)$ transitions.
In all the plots for the observables, we have presented the muon and tau modes only. The reader may note that the plots for electron mode fully overlaps with that of the muon mode.
The average values of these observables considering resonance contributions for transition corresponding to $B^+ \to \rho^+$, $B^0 \to (\rho^0, \omega)$ and $B_s^0 \to \bar{K}^*(892)^{0}$ channels are also listed in Tab. \ref{tab:obs_BBs}.
Note that in Tab. \ref{tab:obs_BBs} the sum of longitudinal and transverse polarizations are different than 1 as the leptons have finite mass in present calculations as opposed to case of massless leptons where it would be unity. The same has been explicitly studied in  \cite{Matias:2012qz,Matias:2012xw}.
We use the Wilson coefficients obtained at the next-to-leading logarithmic order in our calculation of the observables in the full kinematical region of the momentum transfer squared. At this order only the coefficient $C_9^{\rm eff}$ has an imaginary part.
Since our form factors are real, the optimized observable $P'_6$ is identically zero at this order.
Further, motivated by pQCD \cite{Jin:2020jtu} and LCSR \cite{Kindra:2018ayz} approaches, all the physical observables are also computed in the low $q^2$ bins: [0.1 -- 0.98] GeV$^2$ and [1.1 -- 6] GeV$^2$ corresponding to the dropping the charm resonances for electron and muon channels along with the comparison in Tab. \ref{tab:obs_B_bin} - \ref{tab:obs_Bs_bin}.
It is to be noted that the uncertainties in our computations of branching fractions and other physical observables are arising solely because of uncertainties in the form factors. Higher uncertainties are observed in very low $q^2$ range, whereas the maximum propagated uncertainty in the branching fractions is about 49~\% for all the channel $B^{+(0)} \to \rho^{+(0)} e^+ e^-$ when integrated for the entire $q^2$ range. 

\section{Summary and Conclusion}
\label{sec:conclusion}
In this article, we have reported comprehensive study of rare decays corresponding to $b \to d \ell^+ \ell^-$ within the standard model framework of covariant confined quark model with built-in infrared confinement. We first compute the transition form factors in the whole physical range of momentum transfer ($0 \leq q^2 \leq q^2_{\mathrm{max}}$) and utilised for computations of different physical observables such as branching fractions, forward backward asymmetry, transverse and longitudinal polarizations and angular observables.
The observables are computed considering both nonresonant (without vector resonances) and resonant (with light vector and charm resonances) contributions.
The resonance contributions are computed by excluding the experimentally vetoed $q^2$ regions.
We have computed the branching fractions considering both the contributions while the other physical observables are from the vector resonance contributions only.
The computed branching fractions $\mathcal{B}(B^+ \to \pi^+ \mu^+\mu^-)$ is a bit lower and $\mathcal{B}(B_s^0\to \bar{K}^*(892)^0\mu^+ \mu^-)$ branching fraction is within the uncertainty predicted by experimental data from LHCb collaboration, while the nonresonant $\mathcal{B}(B^+\to \rho^+ \mu^+\mu^-) $ is in good agreement within the uncertainty predicted in RQM and LCSR approaches.
Our results on nonresonant rare $B^0$ decays are also matching well with LCSR results for most of the channels.
We also provide the ratio of the branching fractions corresponding to the $b \to s \ell^+ \ell^-$ and $b \to d \ell^+ \ell^-$ utilising the results from previous papers and found that our predictions are consistent with experimental data.
The only exception is the result for ${\mathcal{B} (B^+ \to \pi^+ \mu^+ \mu^-)} / {\mathcal{B} (B^+ \to K^+ \mu^+ \mu^-)}$ in which the ratio is smaller than the measured values due to the small numbers for ${\mathcal{B} (B^+ \to K^+ \mu^+ \mu^-)}$ in our model.
Further, other observables are computed in the whole $q^2$ range as well as in the low $q^2$ range and compared with the LCSR and pQCD approach.
The experimental data on these observables are yet to be reported. We expect these could be measured by LHCb and other $B$ factories as the upgrade II dataset promises the abundance of data for the transition corresponding to $b \to d$ decays.

\section*{ACKNOWLEDGEMENTS}
We would like to thank Prof. Mikhail A. Ivanov for useful discussions of some aspects of this work.
J.N.P. acknowledges financial support from University Grants Commission of India under Major Research Project F.No. 42-775/2013 (SR) and DST-FIST (SR/FST/PS-II/2017/20).
N.R.S. and A.N.G.  thank Bogoliubov Laboratory of Theoretical Physics, Joint Institute for Nuclear Research for warm hospitality during Helmholtz-DIAS International Summer School ``Quantum Field Theory at the Limits: from Strong Field to Heavy Quarks” where work in direction of weak decays initiated. 
This research has been funded by the Science Committee of the Ministry of Education and Science of the Republic of Kazakhstan (Grant No. AP09057862).

\bibliography{apssamp}

\end{document}